\documentclass[aps,nofootinbib,notitlepage,longbibliography,twocolumn,superscriptaddress]{revtex4-1}
\usepackage{amsmath,amssymb,amsfonts}
\usepackage{xcolor}
\usepackage{slashed}
\usepackage{graphicx}
\usepackage{tikz}
\usepackage{bm}
\usepackage{float}
\usepackage[T1]{fontenc}
\usepackage{multirow}
\usepackage[utf8]{inputenc}
\usepackage[colorlinks,linkcolor=blue,citecolor=blue,urlcolor=blue]{hyperref}
\usepackage{subcaption}
\usepackage{ragged2e}
\usepackage{tensor}
\usepackage[compatibility=false]{caption} 
\DeclareCaptionJustification{justified}{\justifying}
\usepackage{booktabs}
\newcommand{\be}{\begin{equation}}
\newcommand{\ee}{\end{equation}}
\newcommand{\bea}{\begin{eqnarray}}
\newcommand{\eea}{\end{eqnarray}}
\newcommand{\ba}{\begin{eqnarray}}
\newcommand{\ea}{\end{eqnarray}}

\newcommand{\ket}[1]{\left|#1\right\rangle}

\def\be{\begin{eqnarray}}
\def\ee{\end{eqnarray}}
\def\bea{\be}
\def\eea{\ee}

\def\roughly#1{\mathrel{\raise.3ex\hbox{$#1$\kern-.75em%
\lower1ex\hbox{$\sim$}}}}

\usepackage{fancyhdr}

\graphicspath{{figures/}}
\begin{document}

\title{Faddeev description of baryons in two-dimensional QCD with $N_c=3$.\\ I. Chiral spectrum, isospin, and strangeness}
\author{Ismail Zahed}
\email[]{ismail.zahed@stonybrook.edu}
\affiliation{Center for Nuclear Theory, Department of Physics and Astronomy,
Stony Brook University, Stony Brook, New York 11794-3800, USA}
\author{Haocheng Zhang}
\email[]{haocheng.zhang@stonybrook.edu}
\affiliation{Center for Nuclear Theory, Department of Physics and Astronomy,
Stony Brook University, Stony Brook, New York 11794-3800, USA}

\date{\today}

\begin{abstract}
We develop a light-front Faddeev description of baryons in two-dimensional QCD at $N_c=3$, in which an exact coupled-channel representation identifies the quark--diquark picture as a single-channel truncation, and solve it variationally with exact matrix elements. In the chiral limit we find an exactly massless lightest baryon, in agreement with the classic valence solutions \cite{Hansson:1978tv,Webber:1979pq}, DLCQ \cite{Hornbostel:1988fb,Burkardt:1989wy}, and bosonization \cite{Steinhardt:1980ry,Date:1986xe}; the excitation tower reproduces Webber's spectrum state by state and sets the valence gap $1.371\,M^{\ast}_{\rm mes}$. The channel content distinguishes these states: the massless baryon is single-channel, the gapped states are genuinely multi-channel, and all baryon Regge trajectories are of the meson class. The calculation is anchored at both ends of the coupling range: the meson-to-baryon mass ratio agrees with strong-coupling bosonization to $0.3$--$4.3\%$, while the valence masses reproduce the full discretized light cone quantization (DLCQ) masses to $0.04\%$ (meson) and $0.15\%$ (baryon) at weak coupling. Extending to two degenerate flavors, we find the massless baryon to be the $\Delta$-like quartet, while the $N$-like doublet at $0.977\,M^{\ast}_{\rm mes}$ is an exact $50\!:\!50$ superposition of good- and bad-diquark channels. Adding a massive strange quark, we obtain, to our knowledge for the first time at the valence three-quark level, the hyperon spectroscopy of QCD$_2$: $\Sigma$-like states fall below $\Lambda$-like ones, inverting the four-dimensional ordering, the chiral thresholds count two kaon units per strange quark, and the Gell-Mann--Okubo relation is protected in mass squared. With all three quark masses non-degenerate, the proton-like state falls below the neutron-like one, $\Sigma^0$--$\Lambda$ mixing appears, and the Coleman--Glashow relation holds at the percent level.
\end{abstract}

\maketitle
\tableofcontents
\thispagestyle{fancy}
\fancyhead{}
\fancyfoot{}
\renewcommand{\headrulewidth}{0pt}

\section{Introduction}

The relativistic three-body problem is a central nonperturbative challenge in quantum field theory. In QCD, confinement implies that baryons are not weakly bound systems of nearly free constituents, but strongly correlated three-quark states. In four dimensions, this problem is difficult. In two dimensions, however, QCD$_2$ retains linear confinement while simplifying the dynamical content, making it an ideal laboratory for developing and testing nonperturbative many-body methods in a relativistic setting.

The present work develops the Faddeev decomposition \cite{Faddeev:1960su} of the baryon equation in QCD$_2$, both at equal time and on the light front, into a complete and computationally viable framework at $N_c=3$. Two conceptual points organize the analysis. First, the Faddeev equations are not a new dynamical theory but an exact reorganization of the three-body problem into pairwise channels; the associated channel decomposition provides structural information that Fock-space diagonalization does not expose, namely which pair configurations build a given baryon. Second, the quark--diquark picture, widely used in the four-dimensional covariant Faddeev approach to baryons \cite{Eichmann:2016yit}, corresponds to a single-channel truncation of this decomposition, and its accuracy becomes a computable question. Both points become sharp in the chiral limit, where the exactly massless chiral baryon \cite{Hansson:1978tv,Webber:1979pq,Hornbostel:1988fb}, the constant pair mode, must be retained in the basis.

We proceed in several stages. First (Sec.~\ref{sec:faddeev}), we formulate the exact three-body baryon equation on the light front and derive its Faddeev decomposition for identical quarks, making the cyclic permutation structure fully explicit. Second (Sec.~\ref{sec:kernel}), we specify the light-front Coulomb kernel appropriate to QCD$_2$ and fix its normalization by matching to the equal-time string tension. Third (Sec.~\ref{sec:coupled}), we derive an exact coupled-channel representation by diagonalizing the intrinsic pair Hamiltonian, thereby exposing the precise relation between the full three-body problem and quark--diquark truncations.

To obtain explicit solutions (Sec.~\ref{sec:corrected}), we solve the symmetric three-quark problem variationally in the physical simplex inner product, in which the kernel quadratic form is manifestly self-adjoint and positive semidefinite and the massless chiral mode is retained. Throughout this work we restrict ourselves to the valence three-quark sector of the light-front Hamiltonian. Within this sector the chiral spectrum consists of the exact massless ground state, with valence distribution $q(x)=6(1-x)$, together with a finite excitation gap generated by pair excitations and their mixing under cyclic permutations. The excitation tower agrees with the permutation-symmetric states of Webber's solution of the same finite-$N_c$ problem \cite{Webber:1979pq} state by state. Since the valence sector is a Tamm--Dancoff truncation of the full light-front theory, excited-state masses are understood as variational upper bounds to the complete spectrum, whereas the exactly massless chiral baryon is protected and survives in the full theory. The Faddeev channel decomposition computed below shows quantitatively that the massless baryon occupies only the lowest pair channel, whereas the excited states necessarily involve coherent multi-channel mixing, explaining why no single-channel (quark--diquark) approximation can describe them.

We then compare our results with strong-coupling bosonization and with the classic discretized light-cone quantization (DLCQ) studies \cite{Pauli:1985pv,Pauli:1985ps,Hornbostel:1988fb,Burkardt:1989wy}, including an absolute comparison at finite quark mass. Extending the calculation to two degenerate flavors introduces no new dynamics but changes the physical content: Fermi statistics promotes the mixed-symmetry spatial states, unphysical for a single flavor, to the isospin-$1/2$ (nucleon-like) tower, yielding the isospin-resolved spectroscopy and the channel anatomy of the two-dimensional ``good'' and ``bad'' diquarks. A massive strange quark then yields the hyperon sector: kaon-like mesons, the $\Sigma$/$\Lambda$/$\Xi$/$\Omega$-like towers (with the $\Sigma$-like states \emph{below} the $\Lambda$-like ones, inverting the four-dimensional ordering), their chiral thresholds, and a two-dimensional test of the Gell-Mann--Okubo relation. Fully non-degenerate masses then yield the isospin-breaking observables: $p$ below $n$, $\Sigma^0$--$\Lambda$ mixing, equal $\Delta$-quartet spacing in mass squared, and the Coleman--Glashow combination.

Webber's SU(6) identification anticipated the classification \cite{Webber:1979pq}, and a qubit-lattice study found the $I=3/2$ assignment of the lightest baryon \cite{Farrell:2022wyt}. The closest prior work is the strong-coupling bosonization treatment of flavored baryons \cite{Date:1986xe,Frishman:1990uw,Ellis:2005ic,Blas:2007dw}. Interest in QCD$_2$ spectroscopy has been renewed by modern Hamiltonian and conformal-field-theory methods \cite{Dempsey:2021xpf,Delmastro:2021otj,Ambrosino:2023dik} and by quantum-simulation studies that prepare and measure $1+1$-dimensional hadrons directly \cite{Atas:2021ext,Farrell:2022wyt}. A Minkowski-space Bethe--Salpeter treatment of the QCD$_2$ baryon has recently rederived the valence light-front problem solved here as its leading truncation, and has solved it at finite quark mass for a single flavor, in the totally symmetric sector only \cite{Kaur:2026bse}. Sharp finite-$N_c$ benchmarks with explicit wavefunctions are directly useful to these programs.

Finally, Appendix~\ref{app:semiclassics} develops a semiclassical analysis of the light-front Faddeev equation, from the Regge-type growth $M_B^2\sim g^2n$ of the excitation spectrum to the reconciliation of finite-$N_c$ pair counting with Witten's large-$N_c$ scaling. Some of the couplings conventions are discussed in Appendix~\ref{app:conventions}.

\section{Faddeev decomposition}
\label{sec:faddeev}

In this section we derive the Faddeev decomposition of the three-body baryon equation, first in a schematic equal-time setting and then on the light front, and reduce it to a single component for identical quarks.

\subsection{Equal-time form}

We begin with the three-body baryon equation in schematic equal-time form (a potential-model illustration of the Faddeev algebra; the genuine equal-time formulation of QCD$_2$, with its Bogoliubov vacuum structure, is developed in Ref.~\cite{Bars:1977ud}),
\begin{equation}
\left(H_0+V_{12}+V_{23}+V_{31}\right)\Psi=M_B\Psi,
\label{eq:ET3body}
\end{equation}
where $H_0$ is the free Hamiltonian and $V_{ij}$ denotes the interaction between quarks $i$ and $j$. In QCD$_2$ the equal-time interaction takes the linear Coulomb form
\begin{equation}
V_{ij}=\sigma\,|x_i-x_j|,
\qquad
\sigma=\frac{g^2}{3}.
\label{eq:stringtension}
\end{equation}
The Faddeev decomposition \cite{Faddeev:1960su} splits $\Psi$ into three components $\Psi_{ij}\propto G_0V_{ij}\Psi$, each obeying an equation in which one pair interacts while the third constituent spectates; the decomposition is mathematically equivalent to Eq.~\eqref{eq:ET3body}, and its value is organizational. We build it once, directly on the light front.

\subsection{Light-front form}

In the light-front three-Fock sector the same logic applies: the light-front baryon equation has the form
\begin{equation}
\left(M_B^2-K_0\right)\Psi(x_1,x_2,x_3)=\sum_{1\le i<j\le 3}K_{ij}\Psi(x_1,x_2,x_3),
\label{eq:LF3body}
\end{equation}
with free operator
\begin{equation}
K_0=\bar m^2\left(\frac{1}{x_1}+\frac{1}{x_2}+\frac{1}{x_3}\right),
\label{eq:K0}
\end{equation}
where $\bar m$ is the (bare) quark mass entering the light-front kinetic term and $x_i>0$ are the three longitudinal momentum fractions satisfying
\begin{equation}
x_1+x_2+x_3=1.
\label{eq:sumx}
\end{equation}
The light-front resolvent is
\begin{equation}
G_0^{\mathrm{LF}}=\frac{1}{M_B^2-K_0},
\end{equation}
and the light-front Faddeev components are defined by
\begin{eqnarray}
\Psi_{12}&=&G_0^{\mathrm{LF}}K_{12}\Psi,
\nonumber\\
\Psi_{23}&=&G_0^{\mathrm{LF}}K_{23}\Psi,
\nonumber\\
\Psi_{31}&=&G_0^{\mathrm{LF}}K_{31}\Psi.
\end{eqnarray}
(For confining kernels, and in particular at the chiral point $M_B^2=0$, $K_0=0$, where each $K_{ij}\Psi$ vanishes on the ground state, this resolvent construction is formal; the single-component equation derived below is well defined throughout and is what we use.) The full wavefunction again satisfies
\begin{equation}
\Psi=\Psi_{12}+\Psi_{23}+\Psi_{31},
\end{equation}
and substitution into Eq.~\eqref{eq:LF3body} gives
\begin{align}
\left(M_B^2-K_0-K_{12}\right)\Psi_{12} &= K_{12}(\Psi_{23}+\Psi_{31}), \\
\left(M_B^2-K_0-K_{23}\right)\Psi_{23} &= K_{23}(\Psi_{31}+\Psi_{12}), \\
\left(M_B^2-K_0-K_{31}\right)\Psi_{31} &= K_{31}(\Psi_{12}+\Psi_{23}).
\label{eq:LFFaddeevcoupled}
\end{align}

\subsection{Identical quarks}

For three identical quarks the baryon wavefunction is symmetric under permutations. One may therefore define a single Faddeev component
\begin{equation}
\Phi(x_1,x_2,x_3)\equiv \Psi_{12}(x_1,x_2,x_3),
\end{equation}
and reconstruct the full wavefunction by cyclic permutations,
\begin{equation}
\Psi=(1+P+P^2)\Phi.
\label{eq:PsiFromPhi}
\end{equation}
The cyclic permutation operator $P$ acts on the quark labels as
\begin{eqnarray}
P&:&(1,2,3)\mapsto (2,3,1),
\nonumber\\
P^2&:&(1,2,3)\mapsto (3,1,2),
\nonumber\\
P^3&=&1.
\end{eqnarray}
Its action on a function of momentum fractions is therefore
\begin{align}
(P\Phi)(x_1,x_2,x_3) &= \Phi(x_2,x_3,x_1), \\
(P^2\Phi)(x_1,x_2,x_3) &= \Phi(x_3,x_1,x_2).
\label{eq:Pactionx}
\end{align}
Using Eq.~\eqref{eq:PsiFromPhi} in the first of Eqs.~\eqref{eq:LFFaddeevcoupled} gives the single exact light-front Faddeev equation for identical quarks,
\begin{equation}
\left(M_B^2-K_0-K_{12}\right)\Phi=K_{12}(P+P^2)\Phi.
\label{eq:singleFaddeev}
\end{equation}
The total wavefunction is then reconstructed from Eq.~\eqref{eq:PsiFromPhi}.

Equation \eqref{eq:singleFaddeev} is the central dynamical equation for the identical-quark problem of Secs.~III--V: what follows there consists of specifying the kernel, reducing the equation to a tractable form, and solving it in a systematic basis. (Unequal quark masses, taken up in Sec.~\ref{sec:twoflavor}, remove the cyclic reduction, and the calculation returns to the full three-kernel form of Eq.~\eqref{eq:LF3body}.)

\section{Light-front pair kernel}
\label{sec:kernel}

This section specifies the standard light-front Coulomb kernel appropriate to QCD$_2$. In one spatial dimension the equal-time interaction is linear in coordinate space, while the corresponding momentum-space light-front kernel is the principal-value Coulomb kernel familiar from the 't~Hooft equation. For the pair $(12)$ this acts on the relative fraction at fixed spectator momentum.

\subsection{Pair variables}

Introduce the pair and spectator variables
\begin{equation}
s=x_1+x_2=1-x_3,
\qquad
z=\frac{x_1}{x_1+x_2}.
\label{eq:szdef}
\end{equation}
Then
\begin{equation}
x_1=sz,
\qquad
x_2=s(1-z),
\qquad
x_3=1-s,
\label{eq:xinSZ}
\end{equation}
with domain
\begin{equation}
0<s<1,
\qquad
0<z<1.
\end{equation}
In these variables the free operator becomes
\begin{equation}
K_0=\frac{\bar m^2}{s}\left(\frac{1}{z}+\frac{1}{1-z}\right)+\frac{\bar m^2}{1-s}.
\label{eq:K0sz}
\end{equation}

\subsection{Definition of \texorpdfstring{$K_{12}$}{K12}}

The pair kernel acting on a test function $f(s,z)$ is defined by
\begin{equation}
(K_{12}f)(s,z)=-\frac{\kappa}{s}\,\mathrm{PV}\int_0^1dz'\,\frac{f(s,z')-f(s,z)}{(z'-z)^2}.
\label{eq:K12def}
\end{equation}
The factor $1/s$ appears because the intrinsic relative coordinate of the interacting pair is measured inside the pair momentum $s=x_1+x_2$.

The coefficient $\kappa$ is fixed by matching to the equal-time string tension. Since the equal-time interaction is $V_{12}=\sigma |x_1-x_2|$ and the one-dimensional Fourier transform of a linear potential produces the Coulomb denominator $1/q^2$ together with a factor of $2$, the light-front kernel coefficient is related to the string tension by
\begin{equation}
\kappa=2\sigma=\frac{2g^2}{3}.
\label{eq:kappaSigma}
\end{equation}
This is the convention used throughout the rest of the paper.

\subsection{Faddeev equation in \texorpdfstring{$(s,z)$}{(s,z)} variables}

Substituting Eqs.~\eqref{eq:K0sz} and \eqref{eq:K12def} into Eq.~\eqref{eq:singleFaddeev} gives
\begin{widetext}
\begin{multline}
\Bigg[M_B^2-\frac{\bar m^2}{s}\left(\frac{1}{z}+\frac{1}{1-z}\right)-\frac{\bar m^2}{1-s}\Bigg]\Phi(s,z) \\
+\frac{\kappa}{s}\,\mathrm{PV}\int_0^1dz'\,\frac{\Phi(s,z')-\Phi(s,z)}{(z'-z)^2}
=(K_{12}P\Phi)(s,z)+(K_{12}P^2\Phi)(s,z).
\label{eq:Faddeevszraw}
\end{multline}
\end{widetext}
Evaluating the right-hand side requires the action of the permutation operators in $(s,z)$ variables.

\subsection{Permutation operator}

Substituting Eq.~\eqref{eq:xinSZ} into the cyclic maps $P:(x_1,x_2,x_3)\mapsto(x_2,x_3,x_1)$ and $P^2$ expresses the permuted arguments of $\Phi$ in the pair variables of the permuted legs:
\begin{equation}
(P\Phi)(s,z)=\Phi\bigg(1-sz,\frac{s(1-z)}{1-sz}\bigg),
\label{eq:Psz}
\end{equation}
\begin{equation}
(P^2\Phi)(s,z)=\Phi\bigg(1-s(1-z),\frac{1-s}{1-s(1-z)}\bigg).
\label{eq:P2sz}
\end{equation}
The reduced light-front equation is therefore completely explicit
\begin{widetext}
\begin{multline}
\Bigg[M_B^2-\frac{\bar m^2}{s}\left(\frac{1}{z}+\frac{1}{1-z}\right)-\frac{\bar m^2}{1-s}\Bigg]\Phi(s,z) \\
+\frac{\kappa}{s}\,\mathrm{PV}\int_0^1dz'\,\frac{\Phi(s,z')-\Phi(s,z)}{(z'-z)^2}
=-\frac{\kappa}{s}\,\mathrm{PV}\int_0^1dz'\,\frac{(P+P^2)\Phi(s,z')-(P+P^2)\Phi(s,z)}{(z'-z)^2},
\label{eq:explicitFaddeevSZ}
\end{multline}
\end{widetext}
where $P\Phi$ and $P^2\Phi$ are given by Eqs.~\eqref{eq:Psz} and \eqref{eq:P2sz}.

\section{Exact coupled-channel reduction}
\label{sec:coupled}

In this section we derive an exact coupled-channel representation of the light-front problem by diagonalizing the intrinsic pair dynamics first; this makes precise the relation between the exact three-quark equation and the quark--diquark approximation.

\subsection{Intrinsic pair Hamiltonian}

Define the intrinsic pair Hamiltonian $h_{12}$ acting on functions of $z$ by
\begin{align}
(h_{12}\phi)(z)={}&\bar m^2\left(\frac{1}{z}+\frac{1}{1-z}\right)\phi(z)
\nonumber\\
&-\kappa\,\mathrm{PV}\int_0^1dz'\,\frac{\phi(z')-\phi(z)}{(z'-z)^2}.
\label{eq:h12}
\end{align}
Its eigenvalue equation is
\begin{equation}
h_{12}\varphi_n(z)=\epsilon_n^2\,\varphi_n(z).
\label{eq:pairEV}
\end{equation}
The eigenvalues $\epsilon_n^2$ are the mass-squared spectrum of the interacting $(12)$ subsystem in the adopted light-front kernel convention (we reserve $\mu^2$ for three-body eigenvalues). We choose the eigenfunctions orthonormal,
\begin{equation}
\int_0^1dz\,\varphi_n(z)\varphi_m(z)=\delta_{nm}.
\label{eq:pairON}
\end{equation}
The Faddeev component is then expanded as
\begin{equation}
\Phi(s,z)=\sum_{n=0}^{\infty}\chi_n(s)\varphi_n(z).
\label{eq:PhiChannelExp}
\end{equation}
Substituting Eq.~\eqref{eq:PhiChannelExp} into Eq.~\eqref{eq:singleFaddeev}, using Eq.~\eqref{eq:pairEV}, and projecting onto $\varphi_n$ yields the exact channel-coupled light-front system,
\begin{equation}
\biggl[M_B^2-\frac{\epsilon_n^2}{s}-\frac{\bar m^2}{1-s}\biggr]\chi_n(s)=\sum_{m}\int_0^1\!ds'\,\mathcal V_{nm}(s,s')\chi_m(s').
\label{eq:channelCoupledEq}
\end{equation}
Here the nonlocal kernel $\mathcal V_{nm}$ originates entirely from the permutation source,
\begin{equation}
\mathcal V_{nm}(s,s')=\mathcal V^{(P)}_{nm}(s,s')+\mathcal V^{(P^2)}_{nm}(s,s').
\end{equation}

\subsection{Derivation of the channel kernel}

To derive $\mathcal V_{nm}$ explicitly, one inserts Eq.~\eqref{eq:PhiChannelExp} into Eqs.~\eqref{eq:Psz} and \eqref{eq:P2sz}, changes variables so that the arguments are expressed in terms of the new spectator fraction $s'$, and then projects with $\varphi_n(z)$. The resulting kernels can be written in the compact form
\begin{widetext}
\begin{equation}
\mathcal V^{(P)}_{nm}(s,s')=
\frac{1}{s^2}\left[\epsilon_n^2-U\!\left(\frac{1-s'}{s}\right)\right]
\varphi_n\!\left(\frac{1-s'}{s}\right)
\varphi_m\!\left(\frac{s+s'-1}{s'}\right)
\Theta(s+s'-1),
\label{eq:VkernelP}
\end{equation}
\begin{equation}
\mathcal V^{(P^2)}_{nm}(s,s')=
\frac{1}{s^2}\left[\epsilon_n^2-U\!\left(\frac{s+s'-1}{s}\right)\right]
\varphi_n\!\left(\frac{s+s'-1}{s}\right)
\varphi_m\!\left(\frac{1-s}{s'}\right)
\Theta(s+s'-1),
\label{eq:VkernelP2}
\end{equation}
\end{widetext}
with
\begin{equation}
U(z)=\bar m^2\left(\frac{1}{z}+\frac{1}{1-z}\right).
\end{equation}
The theta function appears because the transformed variables must lie in the unit interval.

Equation \eqref{eq:channelCoupledEq}, together with Eqs.~\eqref{eq:VkernelP} and \eqref{eq:VkernelP2}, is the exact one-dimensional coupled-channel form of the light-front baryon equation. No approximation has yet been made beyond the adoption of the kernel convention in Eq.~\eqref{eq:K12def}.

\subsection{Relation to diquarks}
\label{sec:diquarks}

Truncating Eq.~\eqref{eq:channelCoupledEq} to the single channel $n=0$ yields an effective one-dimensional equation for $\chi_0(s)$: this is the quark--diquark (dominant-channel) approximation in precise form.

However, this truncation is delicate in the chiral limit. For $\bar m=0$, the constant pair mode
\begin{equation}
\varphi_0(z)=1
\end{equation}
is annihilated by the Coulomb kernel because the integrand in Eq.~\eqref{eq:h12} vanishes identically. Consequently,
\begin{equation}
\epsilon_0^2=0.
\end{equation}

This massless solution is not an artifact: the fully symmetric constant wavefunction $\Psi=\mathrm{const}$ is annihilated by \emph{each} pair kernel and is therefore an exact $M_B^2=0$ eigenstate of the full three-body equation. It is the Faddeev-language counterpart of the exactly massless chiral baryon established in DLCQ \cite{Hornbostel:1988fb,Hornbostel:1988ne,Burkardt:1989wy}. The same massless ground state is found in the direct solution of the finite-$N_c$ valence (Bars--Durgut) equation \cite{Bars:1976prl,Bars:1976tz,Durgut:1976jr,Hansson:1978tv,Webber:1979pq}, and in strong-coupling bosonization, where the baryon is a soliton whose mass vanishes as $m_q^{N_c/(2N_c-1)}$ in the chiral limit \cite{Steinhardt:1980ry,Date:1986xe,Frishman:1992mr}. What the pair-shape excitations with nonvanishing intrinsic energy provide is the finite mass \emph{gap} of the interacting valence sector above this massless state; it is computed in Sec.~\ref{sec:corrected}. This two-tier structure, an exactly massless lightest state together with a gapped interacting sector, is expected on general grounds: in two-dimensional gauge theory with massless fundamental quarks the massless sector (which carries the baryon number) decouples from the universal massive sector \cite{Kutasov:1994xq}, and the string tension vanishes linearly with the quark mass, so that no confining scale survives at $m_q=0$ to lift the lightest baryon \cite{Gross:1995bp}.

\section{Variational solution on the simplex}
\label{sec:corrected}

Rather than integrate the coupled-channel equations of Sec.~\ref{sec:coupled} directly, we solve the equivalent variational problem for the full symmetric wavefunction, in which self-adjointness and the massless zero mode are manifest; the channel content of Sec.~\ref{sec:coupled} is then read off from the resulting eigenstates (Sec.~\ref{sec:channels}). The projection must be carried out on the \emph{physical} symmetric state and in the \emph{physical} inner product: the configuration space of the momentum fractions is the 2-simplex $x_1+x_2+x_3=1$, whose measure in the pair variables is $dx_1\,dx_2=s\,ds\,dz$.

\subsection{Inner product and variational formulation}
\label{sec:varform}

For a totally symmetric $\Psi=(1+P+P^2)\Phi$ the eigenvalue problem $M_B^2\Psi=(K_0+K_{12}+K_{23}+K_{31})\Psi$ is variational. In the chiral limit ($K_0=0$),
\begin{equation}
M_B^2[\Psi]=\frac{3\,\langle\Psi|K_{12}|\Psi\rangle_s}{\langle\Psi|\Psi\rangle_s},
\label{eq:variational}
\end{equation}
with the simplex inner product and kernel quadratic form
\begin{align}
\langle\Psi|\Psi\rangle_s &= \int_0^1\!ds\int_0^1\!dz\; s\,\Psi^2(s,z),
\label{eq:simplexnorm}\\
\langle\Psi|K_{12}|\Psi\rangle_s &= \frac{\kappa}{2}\int_0^1\!ds\!\int_0^1\!\!\int_0^1\! dz\,dz'\,
\frac{\left[\Psi(s,z)-\Psi(s,z')\right]^2}{(z-z')^2}.
\label{eq:quadform}
\end{align}
The Jacobian of $dx_1\,dx_2=s\,ds\,dz$ cancels the $1/s$ of the pair kernel, so no endpoint factor is needed in the basis; the quadratic form \eqref{eq:quadform} is manifestly non-negative and self-adjoint; and $\Psi=\mathrm{const}$ is its exact zero mode. A convenient free symmetric basis is built from the elementary symmetric polynomials,
\begin{align}
\psi_{ab}&=(3e_2)^a(27e_3)^b,\nonumber\\
e_2&=x_1x_2+x_2x_3+x_3x_1,\qquad e_3=x_1x_2x_3,
\label{eq:symbasis}
\end{align}
supplemented by canonical (L\"owdin) orthogonalization to control near-linear dependence; the constant $\psi_{00}=1$ is \emph{included}.

\subsection{Basis, units, and the meson reference}
\label{sec:varbasis}

The monomial set $\{\psi_{ab},\,a+b\le L\}$ is complete on the space of symmetric polynomials but ill-conditioned, so the overlap is L\"owdin-orthogonalized with threshold $\tau$; all matrix elements are exact polynomial integrals, and $L$ and $\tau$ are the only numerical parameters of the chiral solution. Both invariants are needed: the single-index symmetric basis of Ref.~\cite{Kaur:2026bse}, which assigns the same polynomial order to all three quarks, retains only $O(L)$ of the $O(L^2)$ symmetric directions and does not become complete as the order grows. For the symmetric monomial basis, we choose level $L=14$ ($120$ monomials), L\"owdin threshold $\tau=10^{-11}$ ($30$ retained states), and Gauss--Legendre orders $(N_s,N_z,N_{z'})=(72,68,69)$, which integrate the polynomial matrix elements exactly; the convergence is documented in Table~\ref{tab:conv}.

\begin{table}[t]
\caption{Convergence of the symmetric chiral baryon tower with the truncation level $L$ and the L\"owdin threshold $\tau$. The retained dimension is $25$, $27$, $28$, $30$ for $L=8$--$14$ at $\tau=10^{-11}$, and $23$--$38$ as $\tau$ varies from $10^{-9}$ to $10^{-13}$ at $L=14$. States up to $k\simeq11$ are stable at the $1\%$ level or better ($10^{-3}$ or better for $k\le8$); higher states are basis-limited and are not used.}
\label{tab:conv}
\begin{ruledtabular}
\begin{tabular}{lcccccc}
 & $k=1$ & $2$ & $3$ & $4$ & $5$ & $6$ \\
\hline
$L=8$  & $11.0623$ & $24.2038$ & $29.0417$ & $39.1329$ & $47.3508$ & $54.542$ \\
$L=10$ & $11.0623$ & $24.2038$ & $29.0417$ & $39.1330$ & $47.3508$ & $54.520$ \\
$L=12$ & $11.0623$ & $24.2038$ & $29.0417$ & $39.1330$ & $47.3508$ & $54.520$ \\
$L=14$ & $11.0623$ & $24.2038$ & $29.0417$ & $39.1329$ & $47.3508$ & $54.522$ \\
\hline
$\tau=10^{-9}$  & $11.0624$ & $24.2040$ & $29.0418$ & $39.1336$ & $47.3519$ & $54.713$ \\
$\tau=10^{-13}$ & $11.0623$ & $24.2037$ & $29.0417$ & $39.1326$ & $47.3507$ & $54.519$ \\
\end{tabular}
\end{ruledtabular}
\end{table}

In the figures and tables below we quote eigenvalues in the dimensionless units of Webber \cite{Webber:1979pq},
\begin{equation}
\mu^2 \equiv \frac{3\pi}{8}\,\frac{M^2}{g_W^2},
\label{eq:webberunits}
\end{equation}
with $g_W$ Webber's coupling normalization; these units apply to mesons and baryons alike, so the two spectra can be displayed on a common scale and ratios of slopes and masses are directly physical. The relation of $g_W$ to the string-tension convention of Secs.~II--III ($g_\kappa$) and to the DLCQ normalization ($g$) involves fixed factors of $\sqrt2$ and $\sqrt\pi$, derived from a single Hamiltonian in Appendix~\ref{app:conventions}.

The 't~Hooft meson tower in these units, needed below as a reference, is computed by the same Rayleigh--Ritz method in the free Legendre basis
\begin{equation}
\phi(x)=\sum_{n=0}^{L}c_n\,P_n(2x-1),
\label{eq:mesonbasis}
\end{equation}
with $x$ the quark momentum fraction. The result,
\begin{equation}
\mu^2_{\rm mes} = \{0,\;5.8817,\;14.1429,\;23.0841,\;32.3041,\;\dots\},
\label{eq:mesontower}
\end{equation}
agrees with the values quoted by Webber \cite{Webber:1979pq}.

\subsection{Chiral spectrum: massless baryon and the valence gap}
\label{sec:chiralspectrum}

\begin{table}[t]
\caption{Fully symmetric chiral baryon spectrum in the units of Eq.~\eqref{eq:webberunits}, compared with the $S$ states read from Webber's published spectrum \cite{Webber:1979pq}. Both calculations are Rayleigh--Ritz upper bounds.}
\label{tab:tower}
\begin{ruledtabular}
\begin{tabular}{lccc}
state & this work & Webber & difference \\
\hline
$0S$ & $0$ (exact) & $0$ & --- \\
$2S$ & $11.062$ & $11.06$ & $+2\times10^{-4}$ \\
$4S$ & $24.204$ & $24.21$ & $-2\times10^{-4}$ \\
$3S$ & $29.042$ & $29.05$ & $-3\times10^{-4}$ \\
$6S$ & $39.13$ & $39.7$ & $-1.4\%$ \\
$5S$ & $47.35$ & $48.2$ & $-1.8\%$ \\
\end{tabular}
\end{ruledtabular}
\end{table}

The chiral spectrum is shown in Table~\ref{tab:tower}. Its two central features are as follows.

\emph{(i) The lightest baryon is exactly massless, and the zero mode is computed, not imposed.} The lowest eigenvalue comes out at $\mu_0^2 = -4\times10^{-14}$, with eigenvector equal to the constant mode to better than $10^{-4}$. Nothing in the construction singles out the constant: it is one direction among $30$ in the orthonormalized space, and its selection as the ground state is a genuine variational outcome, realizing in the Faddeev framework the massless chiral baryon of the valence equation \cite{Hansson:1978tv,Webber:1979pq} and of DLCQ \cite{Hornbostel:1988fb,Burkardt:1989wy}. This conclusion is not limited to the valence truncation. The truncation is a Tamm--Dancoff restriction of the full light-front Hamiltonian, so the valence ground state is a variational upper bound on the lightest $B=1$ mass of the full theory, while unitarity bounds $M^2$ from below by zero: $0\le M^2_{\rm full}\le M^2_{\rm valence}=0$, and the masslessness of the chiral baryon is therefore exact in full QCD$_2$. What the truncation does affect is everything \emph{above} zero, which is why the gap and tower below carry the valence-sector label. The valence equation is also the leading order of a quasi-potential light-front projection of the Minkowski-space three-quark Bethe--Salpeter equation; the neglected Fock sectors are the higher orders of that expansion \cite{Kaur:2026bse}.

\emph{(ii) The excitation tower agrees with Webber's symmetric states.} For a single flavor at $N_c=3$ the color wavefunction is the antisymmetric singlet, so the physical spatial wavefunction is totally symmetric: the $S$ states are the physical baryon spectrum, and the mixed and antisymmetric classes of the three-body problem require additional quantum numbers not present here (they are supplied by flavor in Sec.~\ref{sec:twoflavor}). Table~\ref{tab:tower} shows agreement state by state: the first three excitations agree to a few parts in $10^{4}$ (relative), and the upper two lie $1.4$--$1.8\%$ \emph{below} Webber's values. Since both calculations produce variational upper bounds, values that are lower are better converged; the pattern is exactly what a tighter bound should produce.

The first excitation sets the gap of the valence three-quark sector,
\begin{equation}
\mu_1^2 = 11.062
\quad\Longrightarrow\quad
M_{\rm gap} = 1.371\,M^{\ast}_{\rm mes},
\label{eq:gap}
\end{equation}
where $M^{\ast}_{\rm mes}$ is the lightest massive 't~Hooft meson ($\mu^2=5.8817$); the ratio in Eq.~\eqref{eq:gap} is independent of all coupling conventions. The gap is a statement about the valence three-quark sector: in the full chiral theory the spectrum above the massless baryon is continuous, because massless mesons can be added without changing the invariant mass \cite{Hornbostel:1988fb}; the gap is the mass scale of the \emph{interacting} sector in the sense of the massless-sector decoupling of Ref.~\cite{Kutasov:1994xq}, whose valence projection it is. Within the valence problem the value itself is not new --- it is Webber's $2S$ state; what is new is its identification as the valence gap above the massless state, its channel content (Sec.~\ref{sec:channels}), and the convention bridge that expresses it in DLCQ units.

\subsection{Valence quark distributions}
\label{sec:qx}

\begin{figure}[t]
\includegraphics[width=\columnwidth]{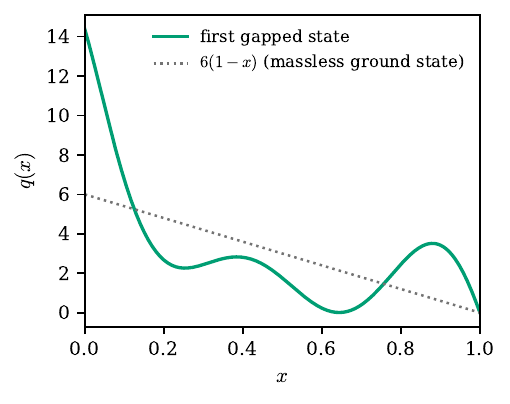}
\caption{Valence quark distribution of the first gapped state, with the massless ground-state distribution $q(x)=6(1-x)$ shown for reference (dotted). The support is concentrated at $x\to0$, the linear-molecule configuration in which one pair carries almost all of the light-front momentum \cite{Webber:1979pq}, with the complementary spectator-pair structure visible as the secondary bump near $x\simeq0.9$.}
\label{fig:qx}
\end{figure}

Each eigenstate carries an explicit wavefunction, and its simplest observable is the single-quark momentum distribution. For a symmetric state $\Psi$ this valence distribution is the simplex marginal
\begin{equation}
q(x_3) = \frac{3}{\mathcal N}\,(1-x_3)\!\int_0^1\! dz\,\big|\Psi(1-x_3,z)\big|^2,
\qquad \int_0^1\! q\,dx = 3,
\label{eq:qxdef}
\end{equation}
with $\mathcal N=\lVert\Psi\rVert_s^2$.
For the constant mode, Eq.~\eqref{eq:qxdef} gives $q(x)=6(1-x)$ \emph{identically}: the linear fall-off is the phase-space factor of the two remaining quarks. This is precisely the continuum distribution of the massless DLCQ baryon, $q(x)=N(N-1)(1-x)^{N-2}$ \cite{Hornbostel:1988fb}. The shape is kinematic rather than dynamical: any light-front space containing the constant mode reproduces it, and the dynamical content is the variational selection of the constant mode as the ground state.

The first gapped state, by contrast, is  dynamical. Its distribution (Fig.~\ref{fig:qx}) rises steeply toward $x\to0$ [$q(0^+)\simeq14$]: one quark is soft while the remaining pair carries almost all of the momentum. This is the linear-molecule structure that Webber identified as dominating the low-lying trajectories \cite{Webber:1979pq}.

\subsection{Faddeev channel content of the chiral states}
\label{sec:channels}

Two distinct channel decompositions arise in the Faddeev framework, and we use both. The coupled-channel representation of Sec.~\ref{sec:coupled} expands the Faddeev \emph{component}, $\Phi=\sum_n\chi_n(s)\varphi_n(z)$, over the eigenmodes $\varphi_n$ of the intrinsic pair Hamiltonian, with intrinsic eigenvalues $\epsilon_n^2/\kappa=\{0,\,5.88,\,14.14,\,\dots\}$ (the 't~Hooft tower). The \emph{physical} pair-mode content of an eigenstate is instead carried by the full symmetric wavefunction $\Psi=(1+P+P^2)\Phi$ in a fixed pairing,
\begin{align}
\Psi(s,z)&=\sum_n \tilde\chi_n(s)\,\varphi_n(z),
\nonumber\\
w_n &= \int_0^1\! s\,ds\,\big|\tilde\chi_n(s)\big|^2\Big/\lVert\Psi\rVert_s^2,
\label{eq:channeldef}
\end{align}
the two being related through the permutation maps of Sec.~\ref{sec:coupled}: the permuted pieces of a single component re-project nontrivially onto the $\varphi_n$, so $\tilde\chi_n\neq\chi_n$ in general. For totally symmetric states the odd channels vanish identically ($\varphi_n(1-z)=(-1)^n\varphi_n(z)$, and exchange symmetry of the pair forces even $z$-parity) so the decomposition runs over $n=0,2,4,\dots$

The component expansion yields an exact statement about the quark--diquark truncation. In the chiral limit the subtracted kernel annihilates the constant in projection, leading to the general result: for \emph{any} $f(s,z)$,
\begin{multline}
\int_0^1\!dz\,\big(K_{12}f\big)(s,z)\\
=-\frac{\kappa}{s}\!\int_0^1\!dz\;\mathrm{PV}\!\!\int_0^1\! dz'\,\frac{f(s,z')-f(s,z)}{(z'-z)^2}=0
\label{eq:constrow}
\end{multline}
by $z\leftrightarrow z'$ antisymmetry. Projecting the component equation \eqref{eq:singleFaddeev} onto $\varphi_0=1$ therefore gives, exactly,
\begin{equation}
M_B^2\,\chi_0(s)=0
\;\Longrightarrow\;
\chi_0\equiv0\ \ \text{for }M_B^2>0:
\label{eq:nogo}
\end{equation}
every gapped eigenstate has identically vanishing $n=0$ component content, and the single-channel ($n=0$) truncation of the coupled equations, the quark--diquark approximation in its precise Faddeev sense, possesses \emph{only} $M_B^2=0$ solutions. The one-channel truncation therefore contains no massive chiral baryon; in particular, it has no counterpart of the gapped state at $\mu^2=11.06$.

The physical pair-mode weights quantify the same physics at the level of the eigenstates themselves:
\begin{itemize}
\item The \emph{massless baryon} is a pure $n=0$ configuration, $w_0=1.000$ to the numerical accuracy of the calculation: all three quarks occupy the lowest pair mode in every pairing, the Faddeev-language expression of the constant wavefunction.
\item The \emph{first gapped state} is genuinely multi-channel: $w_0=0.546$, $w_2=0.452$, $w_4=0.002$. Nearly half of its norm resides in the second pair excitation.
\item The \emph{second gapped state} spreads further: $w_0=0.370$, $w_2=0.361$, $w_4=0.264$ (the remaining $0.5\%$ sits in $n\ge6$).
\end{itemize}
The general result stated above, rules out the single-channel truncation as a dynamical equation; the weights show that it fails even as a \emph{description} of the exact eigenstates: the dominant pair mode carries only $55\%$ of the norm in the first gapped state, and $37\%$ in the second. The gapped baryon sector is a coherent superposition of pair-excited configurations rather than a quark plus a frozen diquark; the one-channel picture is exact only for the massless state, whose diquark is the pair zero mode.

\subsection{Regge trajectories}
\label{sec:reggevar}

\begin{figure}[t]
\includegraphics[width=\columnwidth]{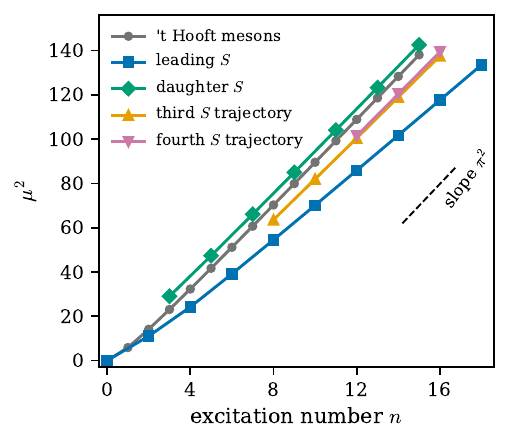}
\caption{Trajectory-resolved chiral spectra versus excitation number $n$, in the units of Eq.~\eqref{eq:webberunits}: 't~Hooft mesons, the leading-$S$ baryon trajectory ($n=0$--$18$), the daughter-$S$ trajectory ($n=3$--$15$), and the third and fourth $S$ trajectories starting at $\mu^2=63.6$ and $101.4$; all members above $\mu^2\simeq48$ lie beyond Webber's published range. The dashed segment indicates the asymptotic meson slope $\pi^2$.}
\label{fig:trajectories}
\end{figure}

Asymptotically the 't~Hooft trajectory is linear, $\mu^2_n\to\pi^2 n$, and Webber observed that the baryon trajectories approach the same slope, differing only in intercept \cite{Webber:1979pq,Hansson:1978tv}. Figure~\ref{fig:trajectories} plots the computed $S$ states, which reach $\mu^2\simeq142$, against the excitation number $n$, the integer of the state labels in Table~\ref{tab:tower}. These states use shifted-Legendre products $P_i(2x_1{-}1)P_j(2x_2{-}1)$ with $i+j\le42$, better conditioned at high excitation than the symmetric basis \eqref{eq:symbasis}, and are identified by their permutation symmetry. Successive members of a trajectory sit at $\Delta n=2$, and everything above $\mu^2\simeq48$ lies beyond Webber's published range. Four trajectories are present: the leading one starts at the massless baryon and occupies even $n$, the daughter starts at $3S$ and occupies odd $n$, and a third and a fourth start at $\mu^2=63.58$ and $101.41$, states that fit neither of the first two. Webber's labels anchor the absolute $n$ only for the first two trajectories. For the third and fourth, the slopes fix the spacing $\Delta n=2$; their horizontal placement in Fig.~\ref{fig:trajectories} is a convention with no physical content, since slopes involve only differences in $n$.

The slopes separate the trajectories into two classes. The daughter, third, and fourth trajectories run parallel to the meson trajectory, with local slopes within $1$--$5\%$ of the meson's at the same $\mu^2$. The leading trajectory rises more slowly: its local slope stays near $0.82$ of the meson's through $n=16$, and its approach to the common asymptote lies beyond the computed range. The channel decomposition of Sec.~\ref{sec:channels} identifies the origin of the two behaviors. A daughter-, third-, or fourth-trajectory state carries most of its norm in a single excited pair mode: in the lowest state of each trajectory the pair predominantly occupies the first, second, and third excited even mode, respectively ($w_2=0.63$, $w_4=0.56$, $w_6=0.43$, with the remainder mostly in the adjacent modes). These are quark--diquark-like configurations: exciting the state along its trajectory stretches the string between the third quark and the pair, and since the pair is a color-$\bar 3$ source, like an antiquark, this string carries the meson tension. Successive trajectories differ only in the internal state of the diquark, so their sequence is the diquark's internal excitation ladder. A leading-trajectory state instead spreads over many pair modes, its largest weight falling from $0.55$ ($n=2$) to $0.24$ ($n=16$).

\subsection{Comparison with bosonization and DLCQ}
\label{sec:quantcomp}

\emph{Bosonization.} Strong-coupling bosonization predicts that the lightest meson and baryon masses vanish together as $m_q\to0$ with the fixed ratio $M_{\rm mes}/M_{\rm bar}\to 2\sin[\pi/(2(2N_c-1))] = 0.618$ \cite{Date:1986xe,Steinhardt:1980ry}; Ref.~\cite{Hornbostel:1988fb} reports consistency with this ratio ``to within the roughly 10\% error'' of their extrapolations (their $m/g=0.2$ point, converted as in Appendix~\ref{app:conventions}, gives $0.70\pm0.12$). In the present framework the massive baryon is obtained by adding the light-front mass term $\beta\sum_i x_i^{-1}$ to the variational calculation, with $\beta$ the dimensionless mass parameter of Appendix~\ref{app:conventions}, and multiplying the basis \eqref{eq:symbasis} by the envelope $(27e_3)^{\delta}$, which carries the non-polynomial endpoint behavior $\Psi\sim x_i^{\delta}$; the exponent is varied about its 't~Hooft endpoint value. Each $\delta$ yields a strict Rayleigh--Ritz upper bound on $M_{\rm bar}$, hence a lower bound on the ratio.

The masses computed in this way are tabulated, together with the strange-quark spectra, in Table~\ref{tab:strange} of Sec.~\ref{sec:hyperons}: the $s\bar s$ and $\Omega$ ($\equiv sss$) columns at $m_s\equiv m$ are the meson and baryon of the present single-flavor problem. In the ratio $M_{\rm mes}/M_{\rm bar}=(\mu^2_{s\bar s}/\mu^2_{\Omega})^{1/2}$ all unit factors cancel [Eq.~\eqref{eq:webberunits}], and the resulting values are
\begin{equation}
\frac{M_{\rm mes}}{M_{\rm bar}} = 0.616,\quad 0.602(1),\quad 0.591(3)
\label{eq:bosonratio}
\end{equation}
at $m/g=0.2$, $0.1$, and $0.05$, to be compared with $0.618$. The pattern of the deviations is physical. The valence truncation omits higher Fock states and therefore overestimates both masses; since the baryon is expected to carry the larger higher-Fock admixture \cite{Hornbostel:1988fb}, the computed ratio should fall below the full-theory prediction, by an amount that grows toward the chiral limit, where the admixture grows. This is the observed pattern. Conversely, $0.618$ is the $m\to0$ limit of the prediction, so at $m/g=0.2$ it carries $O(m/g)$ corrections of its own, and the sub-percent agreement there is partly fortuitous.

\emph{DLCQ: the absolute finite-mass comparison.} The chiral comparisons with DLCQ --- the exactly massless baryon (Sec.~\ref{sec:chiralspectrum}) and the valence distribution $q(x)=6(1-x)$ (Sec.~\ref{sec:qx}) --- are free of unit conventions. The absolute comparison at finite quark mass (Table~\ref{tab:dlcq}) shows the valence masses lying \emph{above} the full DLCQ masses everywhere, by $0.04\%$ (meson) and $0.15\%$ (baryon) at $m/g=1.6$, growing monotonically toward strong coupling as the neglected higher Fock components grow. A Minkowski-space Bethe--Salpeter solution of the same valence equation
\cite{Kaur:2026bse} agrees closely with the valence masses of
Table~\ref{tab:dlcq} once its coupling is normalized to the quark-pair
tension $c_{qq}=C_F/2$ rather than the meson value $C_F$
(Appendix~\ref{app:conventions}). Table~\ref{tab:dlcq}
gives the actual size of the sea effect: the converged valence baryon exceeds
full DLCQ by $0.9\%$ at $m/g=0.8$ and $4.5\%$ at $0.4$.

\begin{table}[t]
\caption{Valence (this work) and full DLCQ \cite{Hornbostel:1988fb} masses at $N_c=3$, the latter converted from program units by $M/g=\sqrt{T\,(m^2/g^2+1/\pi)}$ (Appendix~\ref{app:conventions}), with uncertainties from the quoted extrapolation errors; baryon entries below $m/g=0.4$ are omitted (large DLCQ errors).}
\label{tab:dlcq}
\begin{ruledtabular}
\begin{tabular}{lcccc}
 & \multicolumn{2}{c}{$M_{\rm mes}/g$} & \multicolumn{2}{c}{$M_{\rm bar}/g$} \\
$m/g$ & valence & DLCQ & valence & DLCQ \\
\hline
$1.6$ & $3.647$ & $3.646(2)$ & $5.560$ & $5.552(5)$ \\
$0.8$ & $2.057$ & $2.053(12)$ & $3.185$ & $3.157(15)$ \\
$0.4$ & $1.235$ & $1.218(98)$ & $1.954$ & $1.87(8)$ \\
$0.2$ & $0.785$ & $0.73(12)$ & $1.274$ & \\
$0.1$ & $0.522(1)$ & $0.41(8)$ & $0.867(3)$ & \\
$0.05$ & $0.357(4)$ & $0.25(13)$ & $0.603(19)$ & \\
\end{tabular}
\end{ruledtabular}
\end{table}

\section{Flavor: isospin and strangeness}
\label{sec:twoflavor}

The kernel does not depend on flavor. Adding flavors therefore changes only two inputs: the set of spatial wavefunctions admitted by Fermi statistics, and one mass parameter per massive flavor. This section treats three cases: two degenerate light flavors (isospin), where the mixed-symmetry states become physical; a third, massive flavor (strangeness), which yields kaon-like mesons and the full hyperon spectroscopy; and three unequal masses, where isospin breaking becomes computable.

\subsection{Two degenerate flavors: isospin}
\label{sec:nf2}

The color wavefunction $\epsilon_{abc}$ is totally antisymmetric, and there is no spin in two dimensions, so Fermi statistics requires the flavor$\,\otimes\,$spatial wavefunction to be totally symmetric. For a single flavor this selects the totally symmetric ($S$) spatial states analyzed above. For $N_f=2$ the three-quark flavor states decompose as $\mathbf2\otimes\mathbf2\otimes\mathbf2=\mathbf4_S\oplus\mathbf2_M\oplus\mathbf2_M$: the $I=3/2$ quartet is flavor-symmetric and pairs with the $S$ spatial states (the tower already computed) while the $I=1/2$ doublet is of mixed flavor symmetry and pairs with the mixed-symmetry ($M$) spatial states, the doubly degenerate two-dimensional representation of $S_3$, which is unphysical at $N_f=1$. Totally antisymmetric ($A$) spatial states require an antisymmetric three-quark flavor state and first become physical at $N_f\ge3$. This is the two-dimensional remnant of the familiar $\mathbf{56}/\mathbf{70}/\mathbf{20}$ classification, as Webber already noted in the SU(6) language \cite{Webber:1979pq}. A single-flavor calculation is therefore confined to the symmetric sector, and once flavor is attached its tower carries $I=3/2$: the finite-mass spectrum of Ref.~\cite{Kaur:2026bse}, compared there with the experimental $N(939)$--$N(2300)$ states, is by this assignment $\Delta$-like, and nucleon-like states are outside the reach of a totally symmetric basis.

\begin{figure}[t]
\includegraphics[width=\columnwidth]{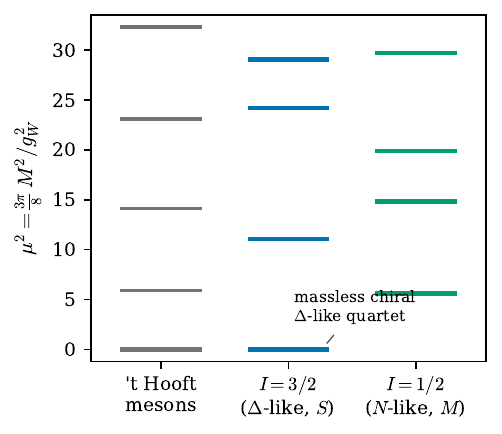}
\caption{Isospin-resolved chiral towers in the units of Eq.~\eqref{eq:webberunits}: 't~Hooft mesons, the $I=3/2$ ($\Delta$-like) tower built on the $S$ spatial states (anchored against Webber in Table~\ref{tab:tower}), and the $I=1/2$ ($N$-like) tower built on the mixed-symmetry states.}
\label{fig:twoflavor}
\end{figure}

To compute the mixed-symmetry tower we repeat the variational calculation of Sec.~\ref{sec:varform} in the \emph{full} (symmetry-unrestricted) polynomial space on the simplex,
\begin{equation}
\Psi(x_1,x_2)=\sum_{i+j\le D}\! c_{ij}\,P_i(2x_1{-}1)\,P_j(2x_2{-}1),\quad D=14,
\label{eq:legprod}
\end{equation}
with the kernel now assembled as $K_{12}+K_{23}+K_{31}$. The $S_3$ classification emerges without being imposed, the $M$ levels appearing as degenerate pairs. The resulting $I=1/2$ tower,
\begin{equation}
\mu^2_{N} = \{5.613,\;14.820,\;19.848,\;29.69,\dots\},
\label{eq:nucleontower}
\end{equation}
reproduces Webber's $M$ states \cite{Webber:1979pq} state by state, now with a physical interpretation they lacked at $N_f=1$.

Fig.~\ref{fig:twoflavor} exhibits two features. First, the chiral hierarchy is \emph{inverted} relative to four-dimensional expectations: the massless chiral baryon is the $I=3/2$ ($\Delta$-like) multiplet (as also found in the qubit-lattice study of Ref.~\cite{Farrell:2022wyt}), because only the totally symmetric spatial sector contains the constant mode, while the $N$-like doublet is gapped at $M_N=0.977\,M^{\ast}_{\rm mes}$. The gap, like that of Eq.~\eqref{eq:gap}, is a valence-sector statement (Sec.~\ref{sec:chiralspectrum}). In the strong-coupling (bosonization) limit the multiplet masses vanish with a common fractional power of the quark mass, $m_q^{1/(1+p)}$ with $p=(N_c^2-1)/N_c(N_c+N_f)$ ($\simeq m_q^{0.65}$ here), and only the coefficients distinguish the multiplets \cite{Date:1986xe}. The same limit also re-inverts the ordering: the flavor-symmetric multiplet becomes the \emph{heaviest}, with a decuplet-to-octet mass ratio of $1.41$ \cite{Date:1986xe}. The hierarchy computed here is therefore specific to the valence sector. Within it the ordering $\Delta(0) < N(0.977\,M^{\ast}_{\rm mes}) < \rho\text{-like}(M^{\ast}_{\rm mes}) < \Delta^{\ast}(1.371\,M^{\ast}_{\rm mes})$ is unambiguous, and the $N$-like baryon, lying \emph{below} the lightest massive meson, realizes in isospin-resolved form Webber's observation that many baryons of this model are lighter than the corresponding mesons.

Second, the channel decomposition of Sec.~\ref{sec:channels} extends to the $N$-like state with one new element. A totally symmetric wavefunction is even under exchange of the two pair quarks, so at $N_f=1$ only the even pair modes $n=0,2,\dots$ appear. The mixed-symmetry wavefunction has both an even and an odd part under this exchange, and the odd part opens the odd channels. By Fermi statistics the odd part carries the flavor-antisymmetric $I=0$ ($ud{-}du$) diquark, the two-dimensional counterpart of the ``good'' scalar diquark of four-dimensional phenomenology \cite{Eichmann:2016yit}, while the even part carries the $I=1$ (``bad'') diquark. Resolving the degenerate $M$ pair into eigenstates of the exchange of quarks $1$ and $2$, $\tilde\chi^{\rm MS}$ (even) and $\tilde\chi^{\rm MA}$ (odd), the physical nucleon-like state $\ket{N}=(\ket{\rm MS}_{\rm space}\ket{\rm MS}_{\rm flavor}+\ket{\rm MA}_{\rm space}\ket{\rm MA}_{\rm flavor})/\sqrt2$ carries exactly half its norm in each isospin class of any chosen pair. The dynamics is in the distribution \emph{within} each class, and it is concentrated in single pair modes: the $I=1$ component is $98.6\%$ pair-ground ($n=0$) channel, and the $I=0$ component is $99.95\%$ in the single $n=1$ mode ($\epsilon_1^2/\kappa=5.88$). Since the intrinsic pair spectrum coincides with the 't~Hooft tower (Sec.~\ref{sec:coupled}), this says that the good diquark inside the two-dimensional nucleon is, in its internal structure, the lightest \emph{massive} 't~Hooft meson mode, while the bad diquark is the massless (constant) one. The two-dimensional nucleon is, to excellent accuracy, an equal coherent superposition of ``quark $+$ bad diquark'' and ``quark $+$ good diquark'', each frozen in its lowest allowed internal mode. In this respect the nucleon is far closer to a two-channel quark--diquark system than the $\Delta$-like gapped states, whose weights spread over channels (Sec.~\ref{sec:channels}).

A caution on units. The pair tower $\epsilon_n^2/\kappa$ is normalized to the \emph{pair} string tension $c_{qq}=C_F/2$, half the meson value $C_F=(N_c^2-1)/2N_c=4/3$. In the common units of Eq.~\eqref{eq:webberunits} the $n=1$ intrinsic energy is $(c_{qq}/C_F)\times5.88=2.94$, so the good diquark's internal energy is a large fraction of, but does not exhaust, $\mu^2_N=5.613$; the remainder is spectator kinematics and permutation coupling.

Secs.~\ref{sec:kaon}--\ref{sec:massrel} add a strange quark, keeping the light quarks exactly chiral, $m_u=m_d=0$, so that the entire flavor-broken spectroscopy is a one-parameter family in $m_s/g$.

\subsection{A strange quark: kaon-like mesons}
\label{sec:kaon}

The strange sector begins with the kaon-like meson, which fixes the strange mass scale. The kaon-like state ($m_1=0$, $m_2=m_s$) is solved in the basis $(1-x)^{\delta_s}P_n(2x-1)$, $n\le40$. The exponent $\delta_s$ sits at the massive endpoint and is fixed by the pair boundary condition of Sec.~\ref{sec:quantcomp}, with $\beta_s=\pi m_s^2/(C_F g^2)$. The massless endpoint carries no exponent: at $m_q=0$ the wavefunction need not vanish at the boundary, the same property that made the chiral baryon massless.

The resulting kaon trajectory interpolates the two known limits. At small quark masses the 't~Hooft model obeys a relation of Gell-Mann--Oakes--Renner (GMOR) type \cite{GellMann:1968rz}, $M^2=(m_1+m_2)\sqrt{\pi g^2C_F/3}+O(m^2)$: each quark mass contributes additively \cite{Zhitnitsky:1985um,Artemev:2025cev}. For the kaon only the strange mass contributes,
\begin{equation}
M_K^2 = g\,m_s\sqrt{\tfrac{\pi C_F}{3}}+O(m_s^2),
\label{eq:kaongmor}
\end{equation}
half the chiral slope of the $s\bar s$ meson. Our small-mass slope, fitted with linear and quadratic terms over $m_s/g\le0.2$, reproduces the predicted coefficient to $0.06\%$. The first excited (``$K^\ast$-like'') state descends continuously from the first massive chiral meson, $\mu^2\to5.8817$ as $m_s\to0$. Numerical values along the $m_s$ grid appear in Table~\ref{tab:strange} together with the baryons.

\subsection{Hyperons: strangeness-resolved baryon spectroscopy}
\label{sec:hyperons}

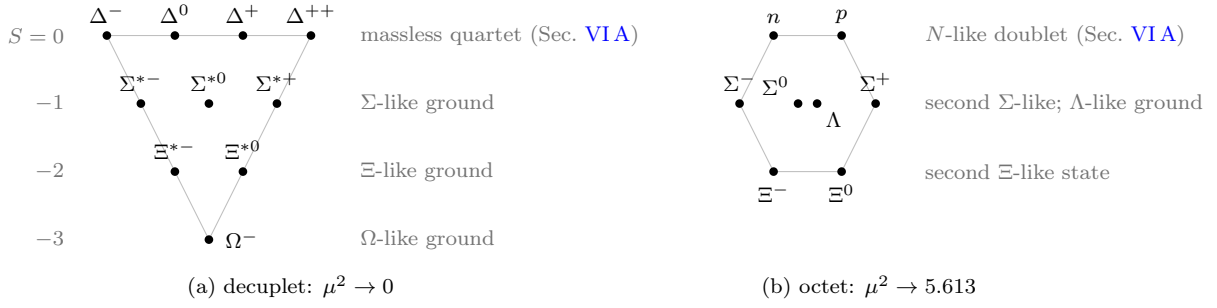
\begin{figure*}[t]
\centering
\begin{tikzpicture}[scale=0.90, every node/.style={font=\footnotesize}]
\begin{scope}
  \draw[black!25] (-1.5,0) -- (1.5,0) -- (0,-3) -- cycle;
  \foreach \x/\lab in {-1.5/$\Delta^-$, -0.5/$\Delta^0$, 0.5/$\Delta^+$, 1.5/$\Delta^{++}$}
    { \fill (\x,0) circle (1.8pt); \node[above=1.5pt] at (\x,0) {\lab}; }
  \foreach \x/\lab in {-1/$\Sigma^{*-}$, 0/$\Sigma^{*0}$, 1/$\Sigma^{*+}$}
    { \fill (\x,-1) circle (1.8pt); \node[above=1.5pt] at (\x,-1) {\lab}; }
  \foreach \x/\lab in {-0.5/$\Xi^{*-}$, 0.5/$\Xi^{*0}$}
    { \fill (\x,-2) circle (1.8pt); \node[above=1.5pt] at (\x,-2) {\lab}; }
  \fill (0,-3) circle (1.8pt); \node[right=3pt] at (0,-3) {$\Omega^-$};
  \node[left, black!55] at (-2.0,0) {$S=0$};
  \node[left, black!55] at (-2.0,-1) {$-1$};
  \node[left, black!55] at (-2.0,-2) {$-2$};
  \node[left, black!55] at (-2.0,-3) {$-3$};
  \node[right, black!55] at (2.1,0) {massless quartet (Sec.~\ref{sec:nf2})};
  \node[right, black!55] at (2.1,-1) {$\Sigma$-like ground};
  \node[right, black!55] at (2.1,-2) {$\Xi$-like ground};
  \node[right, black!55] at (2.1,-3) {$\Omega$-like ground};
  \node at (1.2,-3.7) {(a) decuplet: $\mu^2\to0$};
\end{scope}
\begin{scope}[xshift=8.8cm]
  \draw[black!25] (-0.5,0) -- (0.5,0) -- (1,-1) -- (0.5,-2) -- (-0.5,-2) -- (-1,-1) -- cycle;
  \fill (-0.5,0) circle (1.8pt); \node[above=1.5pt] at (-0.5,0) {$n$};
  \fill (0.5,0) circle (1.8pt); \node[above=1.5pt] at (0.5,0) {$p$};
  \fill (-1,-1) circle (1.8pt); \node[above=1.5pt] at (-1,-1) {$\Sigma^-$};
  \fill (1,-1) circle (1.8pt); \node[above=1.5pt] at (1,-1) {$\Sigma^+$};
  \fill (-0.14,-1) circle (1.8pt); \node[above left=0pt] at (-0.14,-1) {$\Sigma^0$};
  \fill (0.14,-1) circle (1.8pt); \node[below right=0pt] at (0.14,-1) {$\Lambda$};
  \fill (-0.5,-2) circle (1.8pt); \node[below=1.5pt] at (-0.5,-2) {$\Xi^-$};
  \fill (0.5,-2) circle (1.8pt); \node[below=1.5pt] at (0.5,-2) {$\Xi^0$};
  \node[right, black!55] at (1.6,0) {$N$-like doublet (Sec.~\ref{sec:nf2})};
  \node[right, black!55] at (1.6,-1) {second $\Sigma$-like; $\Lambda$-like ground};
  \node[right, black!55] at (1.6,-2) {second $\Xi$-like state};
  \node at (0.9,-3.7) {(b) octet: $\mu^2\to5.613$};
\end{scope}
\end{tikzpicture}
\caption{Two-dimensional analogs of the SU(3) baryon multiplets in the $(I_3,\,\text{strangeness})$ plane as $m_s\to0$. (a) The decuplet, built on the $S$ spatial states, is massless. (b) The octet, built on the mixed-symmetry states, sits at $\mu^2=5.613$.}
\label{fig:multiplets}
\end{figure*}

A strange quark breaks SU(3) flavor to isospin $\times$ strangeness, and the identical-particle constraint now acts separately within each equal-flavor subset. In the one-strange sector ($uus$, $uds$, $dds$) only the two light quarks are identical, and Fermi statistics ties their isospin to their spatial exchange parity as in Sec.~\ref{sec:nf2}: the even tower carries the $\Sigma$-like states ($I=1$), the odd tower the $\Lambda$-like states ($I=0$), and the members of each isospin multiplet are exactly degenerate at $m_u=m_d$. In the two-strange sector ($uss$, $dss$) the $ss$ pair is flavor-symmetric, so only the spatially even $ss$ tower is physical ($\Xi$-like, $I=1/2$, lowest two states denoted $\Xi^\ast$ and $\Xi$); $sss$ is the single-flavor symmetric problem of Sec.~\ref{sec:quantcomp} at $m_q=m_s$ ($\Omega$-like). As $m_s\to0$ flavor SU(3) is restored, and the towers reassemble into its multiplets, $\mathbf3\otimes\mathbf3\otimes\mathbf3=\mathbf{10}\oplus\mathbf8\oplus\mathbf8\oplus\mathbf1$, paired with the $S$, $M$, $A$ spatial classes as in Sec.~\ref{sec:nf2} (Fig.~\ref{fig:multiplets}). The ground states of the $\Sigma$-, $\Xi$-, and $\Omega$-like towers join the massless decuplet ($\Sigma^\ast$, $\Xi^\ast$, $\Omega$). The second $\Sigma$- and $\Xi$-like states and the $\Lambda$-like ground state meet at the octet value $\mu^2=5.613$. The totally antisymmetric state at $25.806$ becomes the flavor singlet, the only one in the computed range; at $m_s>0$ its surviving quantum numbers (one strange quark, $I=0$) are those of the $\Lambda$, so it appears in the spectrum as an excited $\Lambda$-like state. Strong-coupling bosonization has addressed this sector before: flavor multiplets at equal masses \cite{Date:1986xe}, the strangeness content of baryon states \cite{Frishman:1990uw}, exotic-baryon solitons \cite{Ellis:2005ic}, and generalized sine-Gordon solitons at unequal masses with classical mass splittings \cite{Blas:2007dw}. The state-resolved hyperon spectroscopy that follows has, to our knowledge, not been computed.

The computation extends the full-space calculation of Sec.~\ref{sec:nf2}: each strange leg acquires an endpoint envelope,
\begin{equation}
\Psi(x_1,x_2)=\prod_{k\in s}x_k^{\delta_s}\,\sum_{i+j\le D} c_{ij}\,P_i(2x_1{-}1)\,P_j(2x_2{-}1),
\label{eq:hypbasis}
\end{equation}
and the mass term $\beta_s\sum_{k\in s}x_k^{-1}$ is added, with the \emph{pair-normalized} $\beta_s=\pi m_s^2/(c_{qq}g^2)$ (twice the meson-normalized $\beta_s$ of Sec.~\ref{sec:kaon} at $N_c=3$) and $\delta_s$ the corresponding pair boundary-condition exponent, as in Sec.~\ref{sec:quantcomp}. (The physically correct exponent at a strange leg is the \emph{kaon's}: at the endpoint the two strings ending on the strange quark combine into one of meson tension, $2c_{qq}=C_F$ (Sec.~\ref{sec:massrel}). The endpoint analysis of the valence equation gives the same result analytically: the two pair kernels meeting at the endpoint enter the 't~Hooft boundary condition with their strengths added \cite{Kaur:2026bse}. This exponent lies at $0.7$--$0.8$ times the single-pair value, and the variational optimum indeed settles there.) Exchanging the light quarks maps $P_iP_j$ to $P_jP_i$, so the combinations $P_iP_j\pm P_jP_i$ carry definite exchange parity and separate the $\Sigma$-like and $\Lambda$-like towers exactly.

\begin{figure*}[t]
\includegraphics[width=0.95\textwidth]{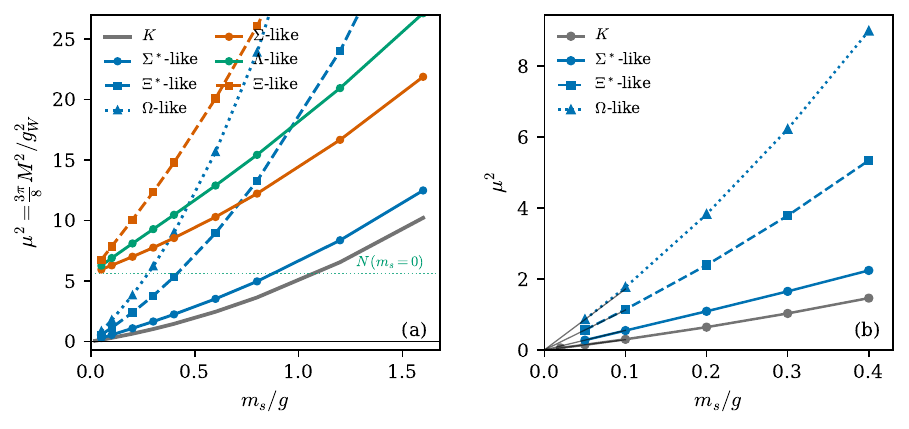}
\caption{Strangeness-resolved spectroscopy, $m_u=m_d=0$. (a) Kaon-like meson and hyperon-like baryon masses squared versus $m_s/g$ in the units of Eq.~\eqref{eq:webberunits}; the $\Lambda$, $\Xi^\ast$, $\Xi$, and $\Omega$ curves leave the frame toward the values of Table~\ref{tab:strange}. The decuplet-like states ($\Sigma^\ast$, $\Xi^\ast$, $\Omega$) rise from the massless chiral baryon; the octet-like $\Sigma$ and $\Lambda$ split from the nucleon value $5.613$. (b) Small-$m_s$ region with the linear (GMOR-like) thresholds indicated; the decuplet-like thresholds approach integer multiples of the kaon threshold as $m_s\to0$, $\mu^2_{\Sigma^\ast}\!:\!\mu^2_{\Xi^\ast}\!:\!\mu^2_{\Omega}\to(2,4,6)\times\mu^2_K$ [Eq.~\eqref{eq:thresholds}].}
\label{fig:strange}
\end{figure*}

\begin{table*}[t]
\caption{Flavor-broken spectroscopy versus $m_s/g$ ($m_u=m_d=0$), in the units of Eq.~\eqref{eq:webberunits}. Mesons: kaon-like ground and first excited state ($K$, $K^\ast$) and the $s\bar s$ ground state; baryons: the lowest two states of the one-strange even tower ($\Sigma^\ast$, $\Sigma$), the lowest odd state ($\Lambda$), the lowest two states of the two-strange even tower ($\Xi^\ast$, $\Xi$), and the $sss$ ground state ($\Omega$). All values are variational upper bounds, except that at $m_s/g\le0.1$ the entries carry the kernel-tail extrapolation, which need not preserve the bound.}
\label{tab:strange}
\begin{ruledtabular}
\begin{tabular}{cccccccccc}
$m_s/g$ & $K$ & $K^\ast$ & $s\bar s$ & $\Sigma^\ast$ & $\Sigma$ & $\Lambda$ & $\Xi^\ast$ & $\Xi$ & $\Omega$ \\
\hline
$0.05$ & $0.145$ & $6.21$ & $0.300$ & $0.276$ & $5.955$ & $6.274$ & $0.563$ & $6.76$ & $0.858$ \\
$0.10$ & $0.300$ & $6.54$ & $0.641$ & $0.546$ & $6.299$ & $6.901$ & $1.142$ & $7.86$ & $1.770$ \\
$0.20$ & $0.643$ & $7.25$ & $1.453$ & $1.089$ & $7.009$ & $8.108$ & $2.390$ & $10.07$ & $3.824$ \\
$0.30$ & $1.029$ & $8.01$ & $2.436$ & $1.651$ & $7.760$ & $9.291$ & $3.785$ & $12.37$ & $6.224$ \\
$0.40$ & $1.461$ & $8.82$ & $3.593$ & $2.241$ & $8.557$ & $10.475$ & $5.341$ & $14.79$ & $8.992$ \\
$0.60$ & $2.459$ & $10.59$ & $6.430$ & $3.526$ & $10.296$ & $12.897$ & $8.957$ & $20.10$ & $15.669$ \\
$0.80$ & $3.641$ & $12.54$ & $9.974$ & $4.969$ & $12.229$ & $15.434$ & $13.268$ & $26.07$ & $23.903$ \\
$1.20$ & $6.559$ & $17.03$ & $19.211$ & $8.370$ & $16.676$ & $20.955$ & $24.031$ & $40.10$ & $45.143$ \\
$1.60$ & $10.222$ & $22.28$ & $31.343$ & $12.491$ & $21.889$ & $27.145$ & $37.695$ & $57.00$ & $72.848$ \\
\end{tabular}
\end{ruledtabular}
\end{table*}

The resulting spectroscopy is shown in Fig.~\ref{fig:strange} and Table~\ref{tab:strange}; as in Sec.~\ref{sec:nf2}, all of it is valence-sector spectroscopy: every splitting quoted here is an observable of the valence truncation. Its principal feature is the $\Sigma$--$\Lambda$ ordering: at every $m_s$ the $\Sigma$-like state lies \emph{below} the $\Lambda$-like state, the reverse of four dimensions, where the color-magnetic interaction favors the $\Lambda$'s scalar diquark. The mechanism is that of Sec.~\ref{sec:nf2}: the $\Lambda$'s good diquark carries the internal energy of the lightest massive 't~Hooft mode, the $\Sigma$'s bad diquark is internally massless, and no spin forces offset the difference. Strangeness adds purity: the diquark picture, only approximate for the nucleon, becomes nearly exact. At $m_s/g=0.4$ the light pair of the $\Lambda$-like ground state carries $99.7\%$ of its norm in the single good-diquark mode, the $\Sigma^\ast$-like state $99.9\%$ and the $\Sigma$-like state $99.6\%$ in the bad-diquark mode. At $m_s=0$ the nucleon mixes the two diquark classes in exactly equal parts, as symmetry requires; a strange quark removes this constraint, and each hyperon settles into a single class. The multi-channel states of Sec.~\ref{sec:channels} reappear only higher in the spectrum. The $\Sigma$--$\Lambda$ splitting opens linearly in $m_s$, like the kaon mass of Eq.~\eqref{eq:kaongmor}, with the $\Lambda$ rising faster ($d\mu^2_\Lambda/d\mu^2_\Sigma=1.93$ at $m_s/g=0.05$); both slopes follow from the chiral wavefunctions alone (Sec.~\ref{sec:massrel}). Toward the heavy-strange limit the mass difference saturates: $\mu_\Lambda-\mu_\Sigma=0.43,\,0.49,\,0.53$ at $m_s/g=0.8,\,1.2,\,1.6$. The saturation says the splitting is a property of the light degrees of freedom alone, as for heavy-quark baryons in four dimensions, but with the opposite sign.

\subsection{Chiral thresholds and mass relations}
\label{sec:massrel}

The small-$m_s$ behavior of Table~\ref{tab:strange} organizes into a simple chiral pattern. Every decuplet-like state rises \emph{linearly} in $m_s$, the baryonic counterpart of the kaon relation \eqref{eq:kaongmor}, and the thresholds are integer multiples of the kaon threshold in the chiral limit:
\begin{align}
\mu^2_{\Sigma^\ast}:\mu^2_{\Xi^\ast}:\mu^2_{\Omega}
&\;=\;\bigl(1.91,\;3.89,\;5.93\bigr)\times\mu^2_K\notag\\
&\;\xrightarrow[m_s\to0]{}\;\bigl(2,\,4,\,6\bigr)\times\mu^2_K,
\label{eq:thresholds}
\end{align}
the measured ratios are quoted at $m_s/g=0.05$ and extrapolate linearly in $m_s$ to the integers within $1\%$. The thresholds follow a boundary-density law:
\begin{equation}
\mu^2_B(m_s)-\mu^2_B(0)\;\to\;\sigma_B\,\mu^2_K(m_s),
\qquad
\sigma_B=\sum_{i\in s}\rho_i(0),
\label{eq:onsetlaw}
\end{equation}
where $\rho_i(x)$ is the momentum-fraction density of strange quark $i$ in the \emph{chiral} ($m_s=0$) state (normalized to $\int\rho_i=1$); for the decuplet-like states $\mu^2_B(0)=0$ and the law fixes the mass itself. The proportionality to $\rho_i(0)$ reflects the mass term $m_s^2/x_s$, which concentrates its weight at the endpoint. The coefficient $\mu^2_K$ reflects the boundary condition there: the pair Couloimb interactions involving the strange quark combine to the meson tension ($2c_{qq}=C_F$; Sec.~\ref{sec:hyperons}), the same as in the kaon, whose own boundary density is $1$. For the constant mode, $\rho(x)=2(1-x)$ (this is $q(x)=6(1-x)$ of Sec.~\ref{sec:qx} per quark) so $\rho(0)=2$ for every strange quark: hence $(2,4,6)$. Consecutive even integers mean equal steps: the decuplet-like tower is equally spaced in mass \emph{squared} in the chiral limit. The successive spacings ($\Delta\to\Sigma^\ast\to\Xi^\ast\to\Omega$) agree to $4\%$ at $m_s/g=0.05$. The four-dimensional decuplet rule is instead equal spacing in the mass $M$ itself, and that version fails here. The difference is the starting point: this decuplet is massless at $m_s=0$, so the breaking is linear in $M^2$, as for mesons in four dimensions, while the four-dimensional decuplet sits on a large SU(3)-symmetric mass, for which the $M$ and $M^2$ rules nearly coincide. The flat meson wavefunction has boundary density $1$ at each end, which makes the $s\bar s$ threshold two kaon units, as the Table confirms ($\mu^2_{s\bar s}/\mu^2_K=2.07$ at $m_s/g=0.05$, extrapolating to $2.01$). The law is not restricted to integer values: for the octet-like states the boundary densities computed from the chiral wavefunctions are $\sigma_\Sigma=2.450$ and $\sigma_\Lambda=4.874$, while the extrapolated thresholds give $2.44$ and $4.83$. The $\Lambda$'s density is the larger because the internal energy of its good diquark enters as $\epsilon_1^2/s$ (Sec.~\ref{sec:coupled}), favoring configurations in which the pair carries most of the momentum and the strange quark is soft; the $\Sigma$'s pair mode is internally massless and exerts no such push.

The octet-like states obey the Gell-Mann--Okubo (GMO) relation \cite{GellMann:1962xb,Okubo:1961jc} in mass squared,
\begin{equation}
2\bigl(\mu^2_N+\mu^2_\Xi\bigr)=3\mu^2_\Lambda+\mu^2_\Sigma,
\label{eq:gmo}
\end{equation}
with $\mu^2_N$ fixed at its chiral value. The defect $\Delta_{\rm GMO}=\tfrac{1}{4}\bigl[2(\mu^2_N+\mu^2_\Xi)-3\mu^2_\Lambda-\mu^2_\Sigma\bigr]$ stays below $0.2\%$ of the mean octet $\mu^2$ for $m_s/g\le0.2$. The textbook derivation of this relation uses first-order perturbation theory in the quark mass, which fails here: on the chiral wavefunctions the matrix element of the mass term diverges logarithmically, and the true shift is proportional to $\sqrt{\beta_s}$, which has no expansion in powers of $\beta_s$. What survives is the structure that $\Delta_{\rm GMO}$ actually tests. The leading shift, Eq.~\eqref{eq:onsetlaw}, acts on one quark at a time and is evaluated in the SU(3)-symmetric states; any one-body flavor operator transforms in $\mathbf3\otimes\bar{\mathbf3}=\mathbf8\oplus\mathbf1$, and octet breaking is precisely the hypothesis from which the GMO relation follows. The non-analytic factor $\sqrt{\beta_s}$ enters only as the overall strength of this operator, and $\Delta_{\rm GMO}$ vanishes for any strength. A nonzero defect requires a $\mathbf{27}$-plet component, which no one-body operator contains. It can come only from the distortion of the wavefunctions by the perturbation, suppressed by a further power of $\sqrt{\beta_s}$. The defect indeed grows roughly linearly in $m_s$, reaching $+4.9\%$ at $m_s/g=0.6$, where the model kaon matches the physical $M_K/M_\rho$ ratio.

The average strange-quark momentum fraction $\langle x_s\rangle$ tests the same diquark dynamics through the bulk of the momentum distribution rather than its endpoint. A quark should carry a growing share of the momentum as it becomes heavy, and the $\Sigma^\ast$ and $\Sigma$ comply: their strange quark carries more than half by $m_s/g=0.4$. In the $\Lambda$ it carries less than an equal third even at these masses. The reason is the push already seen in the boundary densities.

\subsection{Fully non-degenerate masses: isospin breaking}
\label{sec:isobreak}

None of the construction requires any two quark masses to be equal. With $(m_u,m_d,m_s)$ all different, each leg carries its own envelope exponent and mass term, no exchange symmetry survives in the $uds$ channel, and the full space, with all three pair kernels built independently, is diagonalized directly; every spatial state is physical, since no two quarks are identical. We compute at $(m_u,m_d,m_s)=(0.05,\,0.10,\,0.6)\,g$. The ratio $m_d/m_u=2$ is close to its physical value; the absolute light masses are larger than in nature, so that the isospin splittings stand clearly above the numerical uncertainties; and $m_s/g=0.6$ is the value at which the model kaon reproduces the physical $M_K/M_\rho$ ratio (Sec.~\ref{sec:massrel}).

\begin{table}[t]
\caption{Valence spectroscopy at $(m_u,m_d,m_s)=(0.05,0.10,0.6)\,g$, in the units of Eq.~\eqref{eq:webberunits}. The individual masses carry the systematic uncertainties of Table~\ref{tab:strange}; these are nearly common to all states and cancel in the differences.}
\label{tab:isobreak}
\begin{ruledtabular}
\begin{tabular}{lclc}
$\Delta^{++}$ & $0.858$ & $\Sigma^+$ & $11.641$ \\
$\Delta^{+}$  & $1.154$ & $\Sigma^0$ & $12.287$ \\
$\Delta^{0}$  & $1.461$ & $\Sigma^-$ & $12.926$ \\
$\Delta^{-}$  & $1.770$ & $\Lambda$  & $14.358$ \\
$p$           & $7.470$ & $\Xi^0$    & $20.578$ \\
$n$           & $8.233$ & $\Xi^-$    & $21.073$ \\
\end{tabular}
\end{ruledtabular}
\end{table}

The spectrum is collected in Table~\ref{tab:isobreak}. The proton-like state lies below the neutron-like one; this is the sign of nature's splitting, obtained here from quark masses alone (the model contains no electromagnetism). Each isospin multiplet splits, the mass growing with the number of $d$ quarks. The $\Delta$-quartet ($ddd$, $ddu$, $duu$, $uuu$) tests the first-order picture. If the four states shared one spatial wavefunction, each $u\to d$ replacement in the one-body mass term would add the same increment and the quartet would be equally spaced in $\mu^2$. The computed spectrum realizes this spacing closely (Table~\ref{tab:isobreak}): the distortion of the wavefunctions by the light quark masses is small. The Coleman--Glashow combination \cite{Coleman:1961jn}, $(\Xi^--\Xi^0)-(\Sigma^--\Sigma^+)+(n-p)$, is built to vanish at first order in the isospin breaking when the wavefunctions are SU(3) symmetric. Here it comes out at $-2.2\%$ of the $\Sigma$ splitting in $\mu^2$. The residue is proportional to $m_d-m_u$ times the $m_s$-induced distortion of the wavefunctions, the same doubly small cross term that makes the relation accurate in four dimensions.

With $m_u\ne m_d$ the light-pair exchange parity ceases to be a symmetry, and the $\Sigma^0$-like and $\Lambda$-like states mix --- the two-dimensional counterpart of $\Sigma^0$--$\Lambda$ mixing. We measure the admixture spectroscopically through the pair-exchange expectation $\langle P_{12}\rangle=\pm\cos2\theta$ of the two $uds$ eigenstates. The mixing angle is linear in the breaking, $\theta\simeq81^\circ\times(m_d-m_u)/g$ at these masses ($\theta=1.6^\circ,\,4.1^\circ,\,8.0^\circ$ at $m_d-m_u=0.02,\,0.05,\,0.10\,g$, with $\bar m=0.075\,g$ and $m_s=0.6\,g$ fixed). The two $uds$ masses move only at second order in the breaking: the antisymmetric perturbation has no diagonal matrix elements, so it rotates the wavefunctions at first order while leaving the spectrum nearly unchanged, which is why the mixing is measured through $\langle P_{12}\rangle$ rather than through the masses. If only these two states mixed, both would give the same angle; the measured angles differ slightly ($4.1^\circ$ versus $4.9^\circ$ at $m_d-m_u=0.05\,g$), and the difference is the small admixture of higher states, growing smoothly with the breaking. Finally, the boundary-density law \eqref{eq:onsetlaw} is additive across different masses: each quark contributes two kaon units at its own mass, so the lightest ($\Delta$-like) state is predicted at $2\sum_q\mu^2_K(m_q)$. At $(0.05,0.10,0.20)\,g$ the computed mass is $0.956$ of this prediction. The deficit matches the single-flavor thresholds at comparable masses in both sign and size: it is the finite-mass correction to the chiral-limit law, not a failure of additivity.

\section{Conclusion}

We have developed an exact light-front Faddeev formulation of baryons in QCD$_2$ at finite $N_c=3$, and solved its valence three-quark sector using a variational basis on the physical simplex. The Faddeev decomposition provides an exact reorganization of the three-body problem into pair channels, making it possible to quantify the dynamical role of quark--diquark correlations while preserving the complete three-body structure. In the chiral limit the lightest baryon is exactly massless, with the constant pair mode selected variationally rather than imposed, in agreement with the classic valence solutions \cite{Hansson:1978tv,Webber:1979pq}, with DLCQ \cite{Hornbostel:1988fb,Burkardt:1989wy}, and with bosonization \cite{Steinhardt:1980ry,Date:1986xe}; the excitation tower agrees with Webber's permutation-symmetric states state by state and sets the gap of the valence three-quark sector (Table~\ref{tab:tower}).

The channel decomposition is the principal structural result of the formalism: the massless baryon is single-channel, the gapped states divide their norm over several pair channels, and an exact projection identity shows that the quark--diquark truncation possesses no massive chiral solution at all (Sec.~\ref{sec:channels}). The calculation is anchored at both ends of the coupling range: against strong-coupling bosonization, and against the full DLCQ masses, whose long-standing units ambiguity is resolved via the conventions of Ref.~\cite{Hornbostel:1988ne} (Table~\ref{tab:dlcq}).

With two degenerate flavors the same machinery delivers the isospin-resolved spectroscopy of Sec.~\ref{sec:twoflavor}: the massless chiral baryon is the $\Delta$-like quartet, while the $N$-like doublet, an exact superposition of good- and bad-diquark channels, is gapped and slightly below the lightest massive meson. A massive strange quark extends this to the hyperon sector --- $\Sigma$-like states below $\Lambda$-like ones, inverting the four-dimensional ordering; chiral thresholds counting two kaon units per strange quark; Gell--Mann--Okubo protected in mass squared --- and three unequal masses yield the isospin-breaking observables, from the proton--neutron ordering to Coleman--Glashow at the percent level (Secs.~\ref{sec:hyperons}--\ref{sec:isobreak}). To our knowledge, none of this strangeness spectroscopy existed at the valence level before.

The present calculations are restricted to the valence three-quark sector of the light-front Hamiltonian. Consequently, the excited baryon spectrum should be viewed as a controlled variational description of the interacting valence sector, with higher Fock components expected to provide systematic corrections. In contrast, the exactly massless chiral baryon is protected: the valence solution already saturates the lower bound $M_B^2=0$, in agreement with the full DLCQ and bosonized theories \cite{Hornbostel:1988fb,Burkardt:1989wy,Steinhardt:1980ry,Date:1986xe}, so higher Fock sectors cannot shift its mass. The exact Faddeev decomposition itself is independent of this truncation and therefore provides a natural framework for systematically incorporating additional Fock sectors.

The framework extends to channel-truncated solutions of the coupled equations of Sec.~\ref{sec:coupled}, and the explicit wavefunctions and channel weights it produces are directly useful as benchmarks for Hamiltonian, tensor-network, and quantum-simulation studies of QCD$_2$ \cite{Dempsey:2021xpf,Atas:2021ext,Farrell:2022wyt}. More broadly, the two-tier chiral structure quantified here, an exactly massless baryon riding on the decoupled massless sector \cite{Kutasov:1994xq,Delmastro:2021otj} with an $O(g)$ interacting valence sector above it, provides a setting in which confinement, permutation symmetry, and multi-particle correlations can be analyzed in detail.

\begin{acknowledgments}
This work is supported by the U.S. Department of Energy, Office of Science, Office of Nuclear Physics under Contract No. DE-FG-88ER40388. This work is also supported in part by the Quark-Gluon Tomography Topical Collaboration under Award DE-SC0023646.
\end{acknowledgments}

\appendix

\section{Semiclassics of LF Baryons}
\label{app:semiclassics}

In four dimensions, Regge trajectories relate angular momentum and mass squared, $J \sim \alpha' M^2$. In $1+1$ dimensions there is no transverse rotation and hence no angular momentum, but an analogous relation exists between excitation number and mass squared, $M_n^2\sim\alpha'^{-1}n$, where $n$ labels longitudinal excitations.

\subsection{Semiclassical slope estimate}

At high excitation $n\gg1$ the wavefunction oscillates rapidly and a semiclassical treatment applies: channel mixing induced by the permutation operator in Eq.~\eqref{eq:singleFaddeev} becomes subleading, and the dynamics is governed by the intrinsic pair Hamiltonian \eqref{eq:h12} through the diagonal of the coupled-channel equation \eqref{eq:channelCoupledEq},
\begin{equation}
M_B^2 \;\approx\; \frac{\epsilon_n^2}{s} + \frac{\bar m^2}{1-s},
\end{equation}
minimized over the momentum partition $s$; in the chiral limit the pair excitation dominates. The large-$n$ pair spectrum follows from the kernel \eqref{eq:K12def} acting on rapidly oscillating modes $f(z)\sim e^{ikz}$, for which the singular kernel is a pseudodifferential operator with symbol
\begin{equation}
-{\rm PV}\!\int_{-\infty}^{\infty} du\,\frac{e^{iku}-1}{u^2} = \pi |k| ,
\end{equation}
so that, with $k_n\simeq n\pi$ on the unit interval,
\begin{equation}
\epsilon_n^2(s) \simeq \frac{\kappa}{s}\,\pi^2 n ,
\end{equation}
the coupling $\kappa$ being fixed by the string tension [Eqs.~\eqref{eq:stringtension} and \eqref{eq:kappaSigma}]. 

\subsection{Relation to large-$N_c$ baryon scaling}

A baryon Hamiltonian built from three pair kernels is compatible with Witten's large-$N_c$ analysis \cite{Witten:1979kh} (whose group-theoretic content matches the quark model at large $N_c$ \cite{Manohar:1984ys}), in which the baryon mass grows as $M_B\sim N_c$ rather than as the number of quark pairs $N_c(N_c-1)/2$; compatibility requires keeping the color factors and the 't~Hooft scaling distinct.

For two quarks antisymmetrized in color, as in a baryon, the pair color
factor is
\begin{equation}
\langle T_i^a T_j^a\rangle \;=\; -\,\frac{N_c+1}{2N_c}
\;\xrightarrow[N_c\to\infty]{}\; -\frac{1}{2},
\qquad i\neq j,
\end{equation}
of order unity, \emph{not} of order $1/N_c$; correspondingly the pair string
tension $\sigma=g^2(N_c+1)/4N_c$ [$=g^2/3$ at $N_c=3$, Eq.~\eqref{eq:stringtension}]
tends to $g^2/4$. (At $N_c=3$ the numerical coincidence
$(N_c+1)/4N_c=1/N_c$ can obscure this point.) The $1/N_c$ suppression of the
pairwise interaction resides instead in the coupling itself: in the 't~Hooft
limit $g^2=\lambda/N_c$ at fixed $\lambda$ \cite{tHooft:1973alw}. The total pairwise interaction
strength in the baryon then scales as
\begin{equation}
N_{\rm pair}\times\sigma
\;\sim\;\frac{N_c(N_c-1)}{2}\times\frac{\lambda}{4N_c}
\;\sim\;\frac{N_c\lambda}{8},
\end{equation}
linear in $N_c$ at fixed $\lambda$ --- precisely Witten's counting. The
factor of three in the $N_c=3$ saddle estimate is simply $N_{\rm pair}=3$,
the finite-$N_c$ counterpart of the Hartree statement that the baryon
contains $N_c$ coherently interacting constituents; it is not the beginning
of an $N_c^2$ law in 't~Hooft units. Since the realized trajectories are of
linear-molecule type (Sec.~\ref{sec:reggevar}), with a single excited pair,
the asymptotic parallelism of baryon and meson trajectories is a statement
about the pair kernel and is expected to be robust in $N_c$.

\section{Coupling conventions and the mass gap}
\label{app:conventions}
Let $g$ denote the coupling in the normalization of Hornbostel, Brodsky, and Pauli \cite{Hornbostel:1988fb}, in which the light-front meson (valence) equation carries the kernel coefficient $g^2C_F/\pi$ with $C_F=(N_c^2-1)/2N_c=4/3$, and the quark--quark kernel inside the color-antisymmetric baryon carries $g^2c_{qq}/\pi$ with $c_{qq}=(N_c+1)/2N_c=2/3$. Writing the dimensionless 't~Hooft equation as $\mu^2\varphi=\beta\,(x^{-1}+(1-x)^{-1})\varphi+\hat k\varphi$ --- with $\hat k$ the principal-value kernel of Eq.~\eqref{eq:K12def} at unit coefficient and $\bar m$ (written $m$ in ratios such as $m/g$) the bare quark mass of Eq.~\eqref{eq:K0} --- both the mass parameter and the eigenvalue convert with the \emph{same} factor,
\begin{equation}
\frac{\bar m}{g}=\sqrt{\frac{C}{\pi}}\,\sqrt{\beta},
\qquad
\frac{M}{g}=\sqrt{\frac{C}{\pi}}\,\sqrt{\mu^2},
\label{eq:dictionary}
\end{equation}
with $C=C_F$ for the meson and $C=c_{qq}$ per pair for the baryon. Equality of the two factors is required by the free-quark limit: in the free-quark limit $\mu^2\to4\beta$, so $M\to2\bar m$ holds if and only if the factors coincide.

Webber's units \cite{Webber:1979pq}, $\mu^2_W=(3\pi/8)M^2/g_W^2$ for mesons and baryons alike, correspond to Eq.~\eqref{eq:dictionary} with $g_W^2=g^2/2$: his dimensionless spectra are identical to ours, and his coupling differs from the DLCQ-normalized one by $\sqrt2$. The string-tension convention of Secs.~II--III, $\kappa=2\sigma=2g_\kappa^2/3$, matches the physical pair coefficient $g^2c_{qq}/\pi$ for $g_\kappa^2=g^2/\pi$.

For the valence gap of Eq.~\eqref{eq:gap}, $\mu^2_1=11.062$ in Webber units (equivalently a raw pair-normalized eigenvalue $2\mu_1^2=22.125$), the dictionary gives
\begin{equation}
\frac{M_{\rm gap}}{g} = 2.17,
\qquad
\frac{M_{\rm gap}}{g_W} = 3.06,
\qquad
\frac{M_{\rm gap}}{g_\kappa} = 3.84,
\label{eq:gapconventions}
\end{equation}
one number in three normalizations. The only convention-free statements are ratios, of which the most useful are
\begin{align}
\frac{M_{\rm gap}}{M^{\ast}_{\rm mes}}
&=\sqrt{\frac{c_{qq}\,(2\mu^2_1)}{C_F\,\mu^2_{{\rm mes},1}}}
=1.371,
\nonumber\\
\frac{M_{\rm mes}}{M_{\rm bar}}\Big|_{m\to0}&\to
\sqrt{\frac{C_F}{c_{qq}}}\,\sqrt{\frac{\mu^2_{\rm mes}}{2\mu^2_{\rm bar}}},
\label{eq:convfree}
\end{align}
the first evaluated with $\mu^2_{{\rm mes},1}=5.8817$, the second underlying Sec.~\ref{sec:quantcomp} with the fixed prefactor $\sqrt{C_F/c_{qq}}=\sqrt2$.

Finally, the DLCQ mass tables \cite{Hornbostel:1988fb,Hornbostel:1988ne} are given in the units of the DLCQ code, $T=M^2/(m^2+g^2/\pi)$, notwithstanding their $M/g$ column headings; the conversion to physical units is $M/g=\sqrt{T\,(m^2/g^2+1/\pi)}$. This identification follows from the explicit statements of Ref.~\cite{Hornbostel:1988ne} (Secs.~2.2 and 3.1: ``$M^2$ in units of $m^2+g^2/\pi$ are quantities employed directly in the program'') and is confirmed by the axis ranges of the published mass figures.

\bibliography{refs}

@article{Faddeev:1960su,
    author = "Faddeev, L. D.",
    title = "{Scattering Theory for a Three-Particle System}",
    journal = "Sov. Phys. JETP",
    volume = "12",
    pages = "1014--1019",
    year = "1961"
}

@article{tHooft:1973alw,
    author = "'t Hooft, Gerard",
    editor = "Taylor, J. C.",
    title = "{A Planar Diagram Theory for Strong Interactions}",
    reportNumber = "CERN-TH-1786",
    doi = "10.1016/0550-3213(74)90154-0",
    journal = "Nucl. Phys. B",
    volume = "72",
    pages = "461",
    year = "1974"
}

@article{Bars:1976prl,
    author = "Bars, Itzhak",
    title = "{Exact Equivalence of Chromodynamics to a String Theory}",
    reportNumber = "COO-3075-140",
    doi = "10.1103/PhysRevLett.36.1521",
    journal = "Phys. Rev. Lett.",
    volume = "36",
    pages = "1521",
    year = "1976"
}

@article{Bars:1976tz,
    author = "Bars, I.",
    title = "{A Quantum String Theory of Hadrons and Its Relation to Quantum Chromodynamics in Two-Dimensions}",
    reportNumber = "COO-3075-142",
    doi = "10.1016/0550-3213(76)90327-8",
    journal = "Nucl. Phys. B",
    volume = "111",
    pages = "413--440",
    year = "1976"
}

@article{Bars:1977ud,
    author = "Bars, I. and Green, Michael B.",
    title = "{Poincare and Gauge Invariant Two-Dimensional QCD}",
    reportNumber = "COO-3075-179",
    doi = "10.1103/PhysRevD.17.537",
    journal = "Phys. Rev. D",
    volume = "17",
    pages = "537",
    year = "1978"
}

@article{Durgut:1976jr,
    author = "Durgut, Metin",
    title = "{Baryon Bound State in Two-Dimensional SU(N) Gauge Theory}",
    reportNumber = "ITP-SB-76-18",
    doi = "10.1016/0550-3213(76)90324-2",
    journal = "Nucl. Phys. B",
    volume = "116",
    pages = "233--252",
    year = "1976"
}

@article{Hansson:1978tv,
    author = "Hansson, T. H. and Konishi, K.",
    title = "{MULTI - QUARK HADRONS IN TWO-DIMENSIONAL QCD}",
    reportNumber = "RL-78-021",
    month = "2",
    year = "1978"
}

@article{Webber:1979pq,
    author = "Webber, B. R.",
    title = "{Solution of a Two-dimensional {QCD} Model for Baryons}",
    reportNumber = "HEP 79/1",
    doi = "10.1016/0550-3213(79)90609-6",
    journal = "Nucl. Phys. B",
    volume = "153",
    pages = "455--466",
    year = "1979"
}

@article{Manohar:1984ys,
    author = "Manohar, Aneesh V.",
    editor = "Brezin, E. and Wadia, S. R.",
    title = "{Equivalence of the Chiral Soliton and Quark Models in Large N}",
    reportNumber = "HUTP-84/A030",
    doi = "10.1016/0550-3213(84)90584-4",
    journal = "Nucl. Phys. B",
    volume = "248",
    pages = "19",
    year = "1984"
}

@article{Witten:1979kh,
    author = "Witten, Edward",
    title = "{Baryons in the 1/n Expansion}",
    reportNumber = "HUTP-79-A007",
    doi = "10.1016/0550-3213(79)90232-3",
    journal = "Nucl. Phys. B",
    volume = "160",
    pages = "57--115",
    year = "1979"
}

@article{Steinhardt:1980ry,
    author = "Steinhardt, Paul J.",
    title = "{Baryons and Baryonium in {QCD} in Two-dimensions}",
    reportNumber = "HUTP-80/A026",
    doi = "10.1016/0550-3213(80)90065-6",
    journal = "Nucl. Phys. B",
    volume = "176",
    pages = "100--112",
    year = "1980"
}

@article{Date:1986xe,
    author = "Date, Ghanashyam and Frishman, Y. and Sonnenschein, J.",
    title = "{The Spectrum of Multiflavor {QCD} in Two-dimensions}",
    reportNumber = "WIS-86-22-PH",
    doi = "10.1016/0550-3213(87)90278-1",
    journal = "Nucl. Phys. B",
    volume = "283",
    pages = "365--380",
    year = "1987"
}

@mastersthesis{Hornbostel:1988ne,
    author = "Hornbostel, Kent",
    title = "{THE APPLICATION OF LIGHT CONE QUANTIZATION TO QUANTUM CHROMODYNAMICS IN (1+1)-DIMENSIONS}",
    reportNumber = "SLAC-0333, SLAC-333, UMI-89-19433, SLAC-R-0333, SLAC-R-333",
    type = "Ph.D. thesis",
    school = "Stanford University",
    month = "12",
    year = "1988"
}

@article{Burkardt:1989wy,
    author = "Burkardt, M.",
    title = "{The Virial Theorem and the Structure of the Deuteron in (1+1)-dimensional {QCD} on the Light Cone}",
    doi = "10.1016/0375-9474(89)90006-7",
    journal = "Nucl. Phys. A",
    volume = "504",
    pages = "762--776",
    year = "1989"
}

@article{Pauli:1985ps,
    author = "Pauli, Hans Christian and Brodsky, Stanley J.",
    title = "{Discretized Light Cone Quantization: Solution to a Field Theory in One Space One Time Dimensions}",
    reportNumber = "SLAC-PUB-3714",
    doi = "10.1103/PhysRevD.32.2001",
    journal = "Phys. Rev. D",
    volume = "32",
    pages = "2001",
    year = "1985"
}

@article{Pauli:1985pv,
    author = "Pauli, Hans Christian and Brodsky, Stanley J.",
    title = "{Solving Field Theory in One Space One Time Dimension}",
    reportNumber = "SLAC-PUB-3701",
    doi = "10.1103/PhysRevD.32.1993",
    journal = "Phys. Rev. D",
    volume = "32",
    pages = "1993",
    year = "1985"
}

@article{Hornbostel:1988fb,
    author = "Hornbostel, Kent and Brodsky, Stanley J. and Pauli, Hans Christian",
    title = "{Light Cone Quantized QCD in (1+1)-Dimensions}",
    reportNumber = "SLAC-PUB-4678",
    doi = "10.1103/PhysRevD.41.3814",
    journal = "Phys. Rev. D",
    volume = "41",
    pages = "3814",
    year = "1990"
}

@article{Frishman:1992mr,
    author = "Frishman, Y. and Sonnenschein, J.",
    title = "{Bosonization and QCD in two-dimensions}",
    eprint = "hep-th/9207017",
    archivePrefix = "arXiv",
    reportNumber = "WIS-92-54-PH, TAUP-1981-92",
    doi = "10.1016/0370-1573(93)90145-4",
    journal = "Phys. Rept.",
    volume = "223",
    pages = "309--348",
    year = "1993"
}

@article{Eichmann:2016yit,
    author = "Eichmann, Gernot and Sanchis-Alepuz, Helios and Williams, Richard and Alkofer, Reinhard and Fischer, Christian S.",
    title = "{Baryons as relativistic three-quark bound states}",
    eprint = "1606.09602",
    archivePrefix = "arXiv",
    primaryClass = "hep-ph",
    doi = "10.1016/j.ppnp.2016.07.001",
    journal = "Prog. Part. Nucl. Phys.",
    volume = "91",
    pages = "1--100",
    year = "2016"
}

@article{Kutasov:1994xq,
    author = "Kutasov, D. and Schwimmer, A.",
    title = "{Universality in two-dimensional gauge theory}",
    eprint = "hep-th/9501024",
    archivePrefix = "arXiv",
    reportNumber = "EFI-94-67, WIS-12-94",
    doi = "10.1016/0550-3213(95)00106-3",
    journal = "Nucl. Phys. B",
    volume = "442",
    pages = "447--460",
    year = "1995"
}

@article{Gross:1995bp,
    author = "Gross, David J. and Klebanov, Igor R. and Matytsin, Andrei V. and Smilga, Andrei V.",
    title = "{Screening versus confinement in (1+1)-dimensions}",
    eprint = "hep-th/9511104",
    archivePrefix = "arXiv",
    reportNumber = "PUPT-1577, MIT-CTP-2487",
    doi = "10.1016/0550-3213(95)00655-9",
    journal = "Nucl. Phys. B",
    volume = "461",
    pages = "109--130",
    year = "1996"
}

@article{Dempsey:2021xpf,
    author = "Dempsey, Ross and Klebanov, Igor R. and Pufu, Silviu S.",
    title = "{Exact symmetries and threshold states in two-dimensional models for QCD}",
    eprint = "2101.05432",
    archivePrefix = "arXiv",
    primaryClass = "hep-th",
    reportNumber = "PUPT-2623",
    doi = "10.1007/JHEP10(2021)096",
    journal = "JHEP",
    volume = "10",
    pages = "096",
    year = "2021"
}

@article{Delmastro:2021otj,
    author = "Delmastro, Diego and Gomis, Jaume and Yu, Matthew",
    title = "{Infrared phases of 2d QCD}",
    eprint = "2108.02202",
    archivePrefix = "arXiv",
    primaryClass = "hep-th",
    doi = "10.1007/JHEP02(2023)157",
    journal = "JHEP",
    volume = "02",
    pages = "157",
    year = "2023"
}

@article{Ambrosino:2023dik,
    author = "Ambrosino, Federico and Komatsu, Shota",
    title = "{2d QCD and integrability. Part I. {\textquoteright}t Hooft model}",
    eprint = "2312.15598",
    archivePrefix = "arXiv",
    primaryClass = "hep-th",
    reportNumber = "DESY-23-155, CERN-TH-2023-197",
    doi = "10.1007/JHEP02(2025)126",
    journal = "JHEP",
    volume = "02",
    pages = "126",
    year = "2025"
}

@article{Kaur:2026bse,
    author = "Kaur, Satvir and Nair, Sreeraj and Mondal, Chandan and Lan, Jiangshan and Zhao, Xingbo and de Melo, J. P. B. C. and Frederico, Tobias",
    title = "{Baryon Bethe-Salpeter Equation in Minkowski-Space QCD$_2$}",
    eprint = "2605.07095",
    archivePrefix = "arXiv",
    primaryClass = "hep-ph",
    month = "5",
    year = "2026"
}

@article{Ellis:2005ic,
    author = "Ellis, John R. and Frishman, Yitzhak",
    title = "{Exotic baryons in two-dimensional QCD}",
    eprint = "hep-ph/0502193",
    archivePrefix = "arXiv",
    reportNumber = "CERN-PH-TH-2005-017, WIS-22-FEB-DPP",
    doi = "10.1088/1126-6708/2005/08/081",
    journal = "JHEP",
    volume = "08",
    pages = "081",
    year = "2005"
}

@article{Coleman:1961jn,
    author = "Coleman, Sidney R. and Glashow, Sheldon Lee",
    title = "{Electrodynamic properties of baryons in the unitary symmetry scheme}",
    doi = "10.1103/PhysRevLett.6.423",
    journal = "Phys. Rev. Lett.",
    volume = "6",
    pages = "423",
    year = "1961"
}

@article{GellMann:1962xb,
    author = "Gell-Mann, Murray",
    title = "{Symmetries of baryons and mesons}",
    doi = "10.1103/PhysRev.125.1067",
    journal = "Phys. Rev.",
    volume = "125",
    pages = "1067--1084",
    year = "1962"
}

@article{Okubo:1961jc,
    author = "Okubo, Susumu",
    title = "{Note on unitary symmetry in strong interactions}",
    doi = "10.1143/PTP.27.949",
    journal = "Prog. Theor. Phys.",
    volume = "27",
    pages = "949--966",
    year = "1962"
}

@article{GellMann:1968rz,
    author = "Gell-Mann, Murray and Oakes, R. J. and Renner, B.",
    title = "{Behavior of current divergences under SU(3) x SU(3)}",
    doi = "10.1103/PhysRev.175.2195",
    journal = "Phys. Rev.",
    volume = "175",
    pages = "2195--2199",
    year = "1968"
}

@article{Zhitnitsky:1985um,
    author = "Zhitnitsky, A. R.",
    title = "{On Chiral Symmetry Breaking in {QCD} in Two-dimensions ($N_c \to$ Infinity)}",
    reportNumber = "IYF-85-78",
    doi = "10.1016/0370-2693(85)91255-9",
    journal = "Phys. Lett. B",
    volume = "165",
    pages = "405--409",
    year = "1985"
}

@article{Frishman:1990uw,
    author = "Frishman, Y. and Karliner, M.",
    title = "{Baryon Wave Functions and Strangeness Content in {QCD} in Two-dimensions}",
    reportNumber = "WIS-90/13/PH",
    doi = "10.1016/0550-3213(90)90367-M",
    journal = "Nucl. Phys. B",
    volume = "344",
    pages = "393--400",
    year = "1990"
}

@article{Blas:2007dw,
    author = "Blas, Harold",
    title = "{Exotic baryons in two-dimensional QCD and the generalized sine-Gordon solitons}",
    eprint = "hep-th/0702197",
    archivePrefix = "arXiv",
    doi = "10.1088/1126-6708/2007/03/055",
    journal = "JHEP",
    volume = "03",
    pages = "055",
    year = "2007"
}

@article{Artemev:2025cev,
    author = "Artemev, Aleksandr and Litvinov, Alexey and Meshcheriakov, Pavel",
    title = "{QCD2 {\textquoteright}t Hooft model: Two-flavor mesons spectrum}",
    eprint = "2504.12081",
    archivePrefix = "arXiv",
    primaryClass = "hep-th",
    doi = "10.1103/7311-kdbj",
    journal = "Phys. Rev. D",
    volume = "111",
    number = "12",
    pages = "125001",
    year = "2025"
}

@article{Atas:2021ext,
    author = "Atas, Yasar Y. and Zhang, Jinglei and Lewis, Randy and Jahanpour, Amin and Haase, Jan F. and Muschik, Christine A.",
    title = "{SU(2) hadrons on a quantum computer via a variational approach}",
    eprint = "2102.08920",
    archivePrefix = "arXiv",
    primaryClass = "quant-ph",
    doi = "10.1038/s41467-021-26825-4",
    journal = "Nature Commun.",
    volume = "12",
    number = "1",
    pages = "6499",
    year = "2021"
}

@article{Farrell:2022wyt,
    author = "Farrell, Roland C. and Chernyshev, Ivan A. and Powell, Sarah J. M. and Zemlevskiy, Nikita A. and Illa, Marc and Savage, Martin J.",
    title = "{Preparations for quantum simulations of quantum chromodynamics in 1+1 dimensions. I. Axial gauge}",
    eprint = "2207.01731",
    archivePrefix = "arXiv",
    primaryClass = "quant-ph",
    reportNumber = "IQuS@UW-21-027, NT@UW-22-05",
    doi = "10.1103/PhysRevD.107.054512",
    journal = "Phys. Rev. D",
    volume = "107",
    number = "5",
    pages = "054512",
    year = "2023"
}

\end{document}